\documentclass[twocolumn]{aastex62}
\usepackage{graphicx}
\usepackage{bm, amsmath, units}
\usepackage{gensymb}

\graphicspath{{./}{figures/}}

\newcommand*{\Msun}{\ensuremath{\mathrm{M}_\odot}}

\defcitealias{Schneider18a}{Paper 1}
\defcitealias{Schneider18b}{Paper 2}
\defcitealias{Chevalier85}{CC85}

\begin{document}

\title{The Physical Nature of Starburst-Driven Galactic Outflows}

\correspondingauthor{Evan Schneider}
\email{es26@princeton.edu}

\author[0000-0001-9735-7484]{Evan E. Schneider}
\altaffiliation{Hubble Fellow}
\affil{Department of Astrophysical Sciences, Princeton University, 4 Ivy Lane, Princeton, NJ 08540, USA}

\author[0000-0002-0509-9113]{Eve C. Ostriker}
\affil{Department of Astrophysical Sciences, Princeton University, 4 Ivy Lane, Princeton, NJ 08540, USA}

\author[0000-0002-4271-0364]{Brant E. Robertson}
\affiliation{Department of Astronomy and Astrophysics, University of California, Santa Cruz, 1156 High Street, Santa Cruz, CA 95064, USA}

\author[0000-0003-2377-9574]{Todd A. Thompson}
\affiliation{Department of Astronomy and Center for Cosmology and AstroParticle Physics, Ohio State University, 140 W 18th Ave, Columbus, OH, USA}

\begin{abstract}

We present the fourth of the Cholla Galactic OutfLow Simulations suite. Using a physically-motivated prescription for clustered supernova feedback, we successfully drive a multiphase outflow from a disk galaxy. The high resolution ($< 5\,\mathrm{pc}$) across a relatively large domain ($20\,\mathrm{kpc}$) allows us to capture the hydrodynamic mixing and dynamical interactions between the hot and cool ($T \sim 10^4\,\mathrm{K}$) phases in the outflow, which in turn leads to direct evidence of a qualitatively new mechanism for cool gas acceleration in galactic winds. We show that mixing of momentum from the hot phase to the cool phase accelerates the cool gas to $800\,\mathrm{km}\,\mathrm{s}^{-1}$ on kpc scales, with properties inconsistent with the physical models of ram pressure acceleration or with bulk cooling from the hot phase. The mixing process also affects the hot phase, modifying its radial profiles of temperature, density, and velocity from the expectations of radial supersonic flow. This mechanism provides a physical explanation for the high velocity, blue shifted, low ionization absorption lines often observed in the spectra of starburst and high redshift galaxies.

%We present radial profiles and fluxes of relevant physical parameters for both phases, and demonstrate through analytic arguments that the hot phase can be understood as the result of a time-steady flow with a mass source term. We also explore in detail the properties of the cool gas. In particular, we show that the cool  phase can be accelerated to speeds  greater than $800\,\mathrm{km}\,\mathrm{s}^{-1}$ within $10\,\mathrm{kpc}$ of the galaxy, and that the primary mechanism for this acceleration is mixing in momentum from the hot phase. This mechanism provides a robust explanation for the high velocity, blue shifted, low ionization absorption lines often observed in the spectra of starburst and high redshift galaxies.

\end{abstract}


%%%%%%%%%%%%%%%%%%%%%%%%%%%%%%%%%%%%%%%%%%%%%%%%%%%%%%%%%%%%%%%%%%%%%%%

\section{Introduction} \label{sec:introduction}

Theories of galaxy formation now commonly accept that stellar feedback is a necessary ingredient in understanding the way that galaxies evolve \citep[e.g.][and references therein]{Somerville15,Naab17}. On the scale of individual dense star-forming clouds and the surrounding diffuse interstellar medium (ISM), radiation, stellar winds, and supernovae are invoked to explain the low star formation efficiency within clouds and the  low star formation rates in galaxies \citep[e.g.][]{MacLow04,Thompson05,McKee07,Murray10,Ostriker10,Ostriker11,Kim11,Hopkins11,Hopkins14, KKO18,Grudic18,Li19}. On larger scales, feedback in the form of galactic winds and outflows\footnote{In this work, we use the term ``wind" to describe gas that is moving fast enough to escape the galactic halo potential (provided it does not encounter additional surrounding medium), and ``outflow" to describe gas that is moving away from the galaxy at any speed. In other words, all winds are outflows, but not all outflows are winds.} is implicated in the dearth of baryons in galaxies \citep[e.g.][]{Larson74, White78, Dekel86, Keres05, Hopkins14, Genel14, Vogelsberger14, Schaye15, Dave16}, as well as the distribution of metals throughout the circumgalactic medium (CGM) and intergalactic medium (IGM) \citep[e.g.][]{Oppenheimer06, Steidel10,  Peeples11, Dave11, Hummels13, Ford14, Hafen19}. Despite its perceived importance, the details of star-formation driven feedback are less clear. What physical processes drive galactic outflows? How much mass and energy are actually removed from galaxies via star formation feedback? How universal are these properties as a function of galaxy mass and morphology? These are complex questions that require further study.

Observations have shown that outflows are a common feature of star-forming galaxies across a wide range of masses and redshifts \citep[e.g.][]{Martin98, Pettini01, Rubin10, Heckman15,Heckman16,McQuinn19}. Early work using optical spectroscopy found that cool ionized gas can be driven out of galaxies at speeds higher than the escape velocity \citep{Lynds63, Burbidge64}. In low mass galaxies, such as the iconic M82 starburst, the discovery of extended soft X-ray emission \citep{Watson84} led theorists to point to supernovae as the driver of these outflows, positing that an unseen hot ($T \sim 10^{7}$ K) phase existed that could be removing vast quantities of energy from the galaxy in the form of a fast, supersonic wind \citep{Chevalier85}. With the launch of the high resolution Chandra X-ray observatory, this theorized hot plasma was observed directly \citep{Griffiths00, Strickland07}, implicating super-virial gas created by supernovae as a potentially important driver of galactic outflows.

Although models of hot winds explained the process by which metal-rich, supernova heated gas could be driven out of a galaxy, observations of the cooler phases continued to reveal a host of theoretical questions. Taking M82 as an example, in addition to the hot X-ray plasma, outflowing gas has been observed at every wavelength probed, from soft X-ray emission \citep[e.g.]{Strickland04a}, to cool ($T \sim 10^4\,\mathrm{K}$) ionized gas \citep[e.g.][]{McKeith95, Westmoquette09a}, to neutral hydrogen and cold molecular outflows \citep[e.g.]{Walter02, Leroy15, Martini18}. While a fountain flow can explain the decreasing flux of the low velocity molecular gas as a function of height \citep{Leroy15}, a separate mechanism is required to explain the velocities of the faster-moving cool ionized phase, which tend to increase as a function of distance from the galaxy and can exceed the halo escape velocity \citep{Shopbell98}. Down-the-barrel absorption line studies of star-forming galaxies also frequently observe blue-shifted gas in low ionization states, indicating cool outflowing material. This cool ionized gas is observed over a range of velocities, but speeds often reach or exceed $500\,\mathrm{km}\,\mathrm{s}^{-1}$, and some observations see gas moving in excess of $1000\,\mathrm{km}\,\mathrm{s}^{-1}$ 
\citep[e.g][]{Weiner09, Diamond-Stanic12, Martin12, Rubin14, Sell14, Heckman15, Chisholm17}.

Given that the hot gas in winds is theorized to be moving at $v \geq 1000\,\mathrm{km}\,\mathrm{s}^{-1}$, one potential explanation is that the cool phase is simply ISM gas that has been accelerated via ram pressure from the hot gas. A number of idealized studies of cool clouds in hot winds have challenged that explanation, however.   These simulations have demonstrated that ram pressure alone is not effective at accelerating the cool gas, given the competing effects of radiative cooling and subsequent cloud compression that ensue from shocks, and the effects of shear flow interactions on lateral faces. Rather than accelerating clouds, the hot wind tends to heat and destroy them via a combination of shocks and hydrodynamic instabilities \citep[e.g.][]{Nittmann82, Stone92, Klein94, MacLow94, Xu95, Cooper09, Scannapieco15, Bruggen16, Schneider17, Zhang17}. However, a few recent studies have noted that, given large enough clouds and appropriate background conditions, cool gas can persist in these simulations as a result of a mixing and cooling cycle. Under the right circumstances, this may result in an increased flux of cool gas as hot gas condenses out, effectively growing the cloud rather than destroying it \citep{Armillotta16, Gritton17, Gronke18, Gronke19}. Unfortunately, the idealized nature of the background flow in these simulations makes it difficult to tell whether this mechanism is viable in the turbulent, high pressure environment of a galactic wind \citep{Schneider18a, Fielding18}. In an alternative model, \cite{Thompson16} suggested that the hot wind itself could cool to $T \sim 10^{4}\,\mathrm{K}$, provided it was sufficiently mass-loaded via the destruction of cool gas at small radii.

The combination of uncertainties about the physical nature of gas in outflows and the theoretical uncertainties about the mechanisms for accelerating cool gas motivated the Cholla Galactic OutfLow Simulations (CGOLS) project, a series of extremely high resolution global disk simulations of galaxy outflows. By simulating a whole galaxy, the CGOLS project aims to avoid uncertainties related to the limited domain present in cloud-wind or ISM patch simulations, while maintaining high enough resolution to sufficiently capture the hydrodynamic instabilities associated with the destruction of cool gas in winds. In earlier work, we tested the effect of including a central feedback source in a galaxy disk, both with and without radiative cooling, in order to elucidate how well analytic models for supernova driven winds could predict wind properties (\citealt{Schneider18a, Schneider18b}, hereafter \citetalias{Schneider18a} and \citetalias{Schneider18b}). These simulations showed that theoretical wind models work well in scenarios where the hot wind is unaffected by interactions with the gas in the disk, but do not accurately reproduce the properties of the wind in cases where it has experienced significant mass loading, or when the spherical symmetry of the feedback injection scheme is broken. In part, this is because none of the analytic models tested in our earlier work account for the multiphase nature of gas in winds \textit{at a single radius}.

In this work, we present a new CGOLS simulation that includes a multiphase wind, as well as a two-phase analytic model capable of fitting the properties of the wind as a function of radius. The primary difference between this simulation and those presented in \citetalias{Schneider18a} and \citetalias{Schneider18b} is the nature of the feedback injection mechanism, which is described in Section~\ref{sec:cluster_feedback}. We present details of the analytic model in Section~\ref{sec:analytics}. Section~\ref{sec:results} contains the primary results of the simulations, including a discussion of the radial profiles of both the hot and cool phases, as well as radially averaged outflow rates, both of which are components in the analytic model. We also show the mass and energy loading in different phases, as well as phase diagrams demonstrating the relationships between various physical quantities. We then address the mechanism by which the cool gas is accelerated to velocities comparable to those seen in observations, highlighting our method for demonstrating the role of hydrodynamic mixing in the momentum transfer process from hot to cool phases. We conclude the section with a discussion of convergence in our simulations. In Section~\ref{sec:discussion}, we discuss some observational implications of this work, as well as our model dependence, and the relationship between our results and previous simulations presented in the literature. We conclude with a brief summary of our results in Section~\ref{sec:conclusions}.

%%%%%%%%%%%%%%%%%%%%%%%%%%%%%%%%%%%%%%%%%%%%%%%%%%%%%%%%%%%%%%%%%%%%%%%

\section{Simulations} \label{sec:simulations}

Here we briefly describe the overall setup of the simulation - further details of the CGOLS suite can be found in \citetalias{Schneider18a}. Each simulation is carried out in a box with a uniform grid of cells. The box has dimensions $(L_x, L_y, L_z) = (10\,\mathrm{kpc}, 10\,\mathrm{kpc}, 20\,\mathrm{kpc})$, with $(N_\mathrm{cells, x}, N_\mathrm{cells, y}, N_\mathrm{cells, z}) = (2048, 2048, 4096)$, resulting in a constant cell width of $\Delta x = \Delta y = \Delta z = 10\,\mathrm{kpc}/2048 \approx 4.9\,\mathrm{pc}$.

Centered within the box, we place a disk of $10^4\,\mathrm{K}$ isothermal gas, distributed with an exponential surface density profile with scale radius $R_\mathrm{gas} = 1.6\,\mathrm{kpc}$ and total mass $M_\mathrm{gas} = 2.5\times10^9\,\mathrm{M}_\odot$. This corresponds to a central surface density of $\Sigma_0 \approx 150\,\mathrm{M}_\odot\,\mathrm{pc}^{-2}$. The gas disk is initially in vertical hydrostatic and rotational equilibrium with a static gravitational potential composed of a Miyamoto-Nagai profile for the galaxy's stellar disk component \citep{Miyamoto75}, and an NFW profile for the dark matter component \citep{Navarro96}. The disk potential is given by
\begin{equation}
\Phi_\mathrm{stars}(r, z) = - \frac{G M_\mathrm{stars}}{\sqrt{r^2 + \left(R_\mathrm{stars} + \sqrt{z^2 + z_\mathrm{stars}^2}\right)^2}},
\end{equation}
where $r$ and $z$ are the radial and vertical cylindrical coordinates, $M_\mathrm{stars} = 10^{10}\,\mathrm{M}_\odot$ is the mass of the stellar disk, $R_\mathrm{stars} = 0.8\,\mathrm{kpc}$ is the stellar scale radius, and $z_\mathrm{stars} = 0.15\,\mathrm{kpc}$ is the stellar scale height. The values for the gas mass, stellar mass, scale radii, and stellar scale height were set to mimic those of the local starburst galaxy, M82 \citep{Greco12, Mayya09, Lim13}. The dark matter potential is likewise defined by
\begin{equation}
\Phi_\mathrm{halo}(r) = -\frac{G M_\mathrm{halo}}{r[\mathrm{ln}(1 + c) - c/(1+c)]}\mathrm{ln}\left(1+\frac{r}{R_\mathrm{halo}}\right),  
\end{equation}
where $r$ is the radius in spherical coordinates, $M_\mathrm{halo} = 5\times10^{10}\,\mathrm{M}_\odot$ is the assumed dark matter mass of the halo, $c = 10$ is the halo concentration, and $R_\mathrm{halo}$ is the scale radius of the halo, which we set to $R_\mathrm{halo} = R_\mathrm{vir} / c = 5.3\,\mathrm{kpc}$. %These values were also chosen with M82 in mind, though in practice the vertical acceleration is dominated by the stellar disk within the $10\,\mathrm{kpc}$ height of the simulation volume, so the exact values chosen for the dark matter potential are not important. 
Outside of the disk, we place a static hot halo in hydrostatic equilibrium with the potential - this gas is quickly blown away when the simulation starts. 

\subsection{Cluster Feedback}\label{sec:cluster_feedback}

The primary difference between this work and earlier simulations in the CGOLS suite is the implementation of the stellar feedback. In \citetalias{Schneider18b}, we described a method of clustered feedback where mass and energy were deposited in 8 spherical regions within the disk, each with $R = 150\,\mathrm{pc}$. The values of the mass and energy injection were defined arbitrarily such that the volume-filling gas in the resulting hot outflow would either have high enough density to cool efficiently down to $10^4\,\mathrm{K}$ at a relatively small radius (the ``high mass-loading state"), or would not cool efficiently (the ``low mass-loading state"), as described in \cite{Thompson16}. We defined these mass and energy injection rates $M_\mathrm{inj}$ and $E_\mathrm{inj}$ in relation to the star formation rate: mass-loading is quantified by $\beta$, 
\begin{equation}
    \beta = \frac{\dot{M}_\mathrm{inj}}{\dot{M}_\mathrm{SFR}},
\end{equation}
and energy loading is quantified by $\alpha$, such that
\begin{equation}
    \dot{E}_\mathrm{inj} = 3\times10^{41}\,\mathrm{erg}\,\mathrm{s}^{-1}\, \alpha \left[\frac{\dot{M}_\mathrm{SFR}}{M_\odot\,\mathrm{yr}^{-1}}\right],
\end{equation}
where we have assumed that there is 1 supernova per $100\,\mathrm{M}_\odot$ of star formation, and each supernova generates $10^{51}\,\mathrm{erg}$ of energy. In each state, we held the values of $\alpha$ and $\beta$ steady for many millions of years.

In this work, we relax these assumptions about the values of $\alpha$ and $\beta$. Now, we set the mass and energy injection rates in the clusters using physically-motivated values determined by running a separate ``superbubble" simulation. This simulation will be described in detail in a future paper (Schneider \& Ostriker, \textit{in prep}), but in brief - we use a Starburst99 single burst stellar population synthesis model \citep{Leitherer99} to inject mass and energy into a box containing clouds that represent a well-resolved multiphase ISM, and we then track the interaction between phases as the resulting bubble driven by the cluster feedback propagates through the box \cite[see][for an example of such simulations]{Kim17}. The ISM gas that is swept up as the bubble expands before breaking out of the disk can therefore contribute to the mass loading at early times.

\begin{figure}
    \centering
    \includegraphics[width=\linewidth]{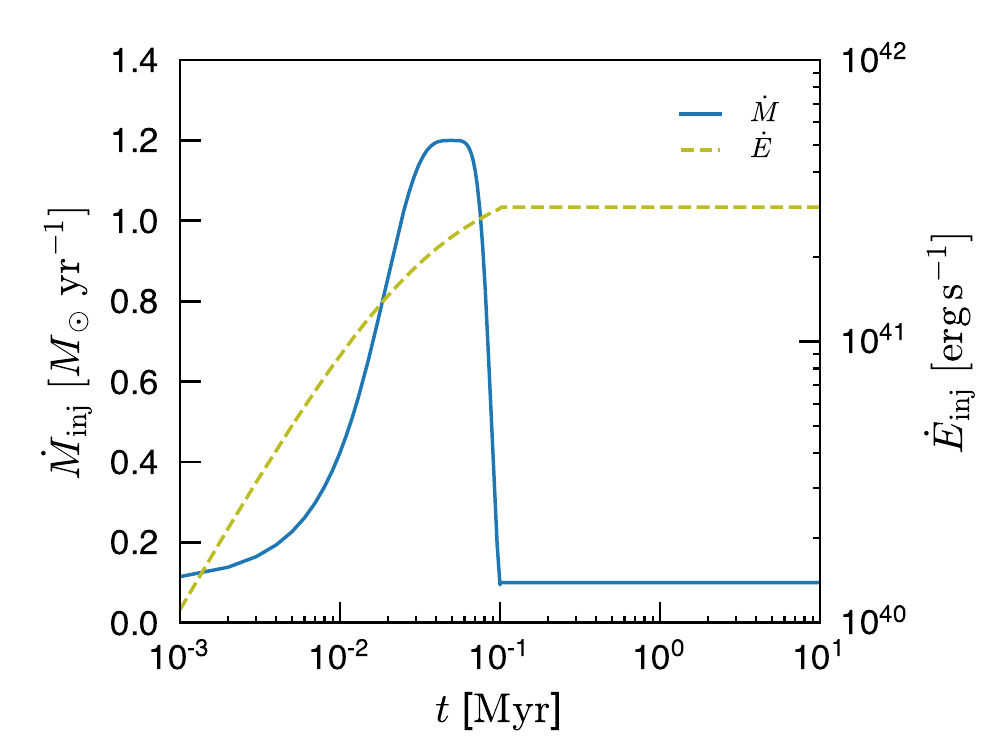}
    \caption{Mass and energy injection rates for each cluster. The blue solid line shows the mass injection rate per cluster (left axis), and the green dashed line shows the energy injection rate (right axis).}
    \label{fig:cluster_inj}
\end{figure}

In order to compare with our earlier, more idealized models, we use very large clusters for the feedback in this simulation - each cluster is assumed to host $10^7\,\Msun$ of stars - and each cluster is turned ``on" for $10\,\mathrm{Myr}$, as in our previous work. For the first $10^5\,\mathrm{yr}$ that the cluster is on, the mass and energy injection rates are defined by the equations
\begin{equation}
  \dot{M}_\mathrm{inj} = 1.2 \times\mathrm{exp}(-\left[\frac{(t-0.05)^{2}}{(2\times 0.028^2)}\right]^2) \,\Msun\,\mathrm{yr}^{-1}
\end{equation}
and
\begin{equation}
   \dot{E}_\mathrm{inj} = 4\times10^{41} \frac{t^2}{(t^2 + 0.035t)}\,\mathrm{erg}\,\mathrm{s}^{-1},
\end{equation}
where $t$ is measured in Myr; thereafter $\dot{M} = 0.1\,\Msun\,\mathrm{yr}^{-1}$ and $\dot{E} = 3\times10^{41}\,\mathrm{erg}\,\mathrm{s}^{-1}$. These mass and energy injection rates are plotted in Figure~\ref{fig:cluster_inj}. 

In terms of an injected $\alpha_\mathrm{inj}$ and $\beta_\mathrm{inj}$ for each cluster, this corresponds to a mass-loading that peaks at $\beta_\mathrm{inj} = 1.2$ as the cluster is breaking out of the disk, accounting for interactions with swept up cold clouds in the multiphase ISM (not present in the CGOLS simulation), then lowers to a value of $\beta_\mathrm{inj} = 0.1$, which is approximately the average for the mass return from supernovae and stellar winds at late times \citep{Leitherer99}. With this model, approximately 10\% of the total mass injection for each cluster occurs during the initial breakout phase. Meanwhile, the energy-loading is a steadily increasing function that reaches $\alpha_\mathrm{inj} = 1$ after $10^5\,\mathrm{yr}$, by which time a cluster of this size will have broken out of the disk, and energy losses within the $R_\mathrm{cl} = 30\,\mathrm{pc}$ injection region should be very low \citep{Fielding18}. Note that these values for $\alpha_\mathrm{inj}$ and $\beta_\mathrm{inj}$ are defined in terms of an assumed ``star formation rate" of $1\Msun\,\mathrm{yr}^{-1}$ in each cluster, defined as the total mass divided by the time the cluster is turned on, $\dot{M}_\mathrm{SFR} = 10^7\Msun / 10^7\mathrm{yr}$. These functions for $\dot{M}$ and $\dot{E}$ are fits to the actual measured $\dot{M}$ and $\dot{E}$ values at a radius corresponding to a cluster in the CGOLS simulation ($R_\mathrm{cl} = 30\,\mathrm{pc}$) as measured in a superbubble simulation with a similar average gas density (Schneider \& Ostriker, \textit{in prep.}). We emphasize that these are the \textit{injected} mass and energy rates within the clusters - the effective $\alpha$ and $\beta$ measured in the wind can be very different as a result of interactions with gas in the disk.

Clusters are turned on every million years at a rate corresponding to $20\,\Msun\,\mathrm{yr}^{-1}$ of star formation from 5-35~Myr, and $5\,\Msun\,\mathrm{yr}^{-1}$ for the remainder of the simulation, which runs to 70 Myr. The clusters are randomly distributed in radius and azimuthal angle throughout the central $R = 1\,\mathrm{kpc}$ of the disk, and up to $z = 5\,\mathrm{pc}$ above or below the midplane. (We note that this 1 kpc radius for cluster distribution is significantly larger than that of the central starburst in M82; despite using it to model our initial conditions, we are not attempting to replicate that system in detail.) After being turned on, each cluster rotates with the disk at a speed set by the gravitational potential. At the end of its 10 Myr lifespan, each cluster is turned off. This cluster lifespan was chosen to match our previous simulations, and we will investigate the effects of a longer cluster lifetime in future work.

While this model represents a rather extreme mode of centralized star cluster feedback, it is not wholly unphysical. In any given starburst, the energetics of the wind will be driven by the most massive clusters. In the CGOLS simulations, we do not resolve the details of the multiphase ISM, and therefore any smaller clusters that would not break out of the disk may be neglected for the purposes of studying the outflow - their primary effect would be to inject momentum into the ISM. On the other hand, including a large number of more moderately-sized ($M_* \sim 10^5 - 10^6\,\Msun$), longer lived clusters would have the effect of significantly increasing the surface area of interaction between clusters and disk gas, and could therefore plausibly increase the mass-loading rate substantially, while lowering the effective value of $\alpha$ in the outflow for a given star formation rate. We will test the effects of these assumptions in future work by examining a simulation with a distribution of cluster masses.

\subsection{The Passive Scalar}
Like the previous CGOLS simulations, the simulation presented in this work was carried out using the Cholla hydrodynamics code \citep{Schneider15}, using a PLM reconstruction scheme, an HLLC Riemann solver, and an unsplit predictor-corrector integration method \citep{Stone09}. We also employ a dual-energy method to track the internal energy of the gas, given the high Mach numbers attained by cool gas in the outflow \citep{Bryan95, Schneider17}. Unlike the previous simulations, in addition to evolving the conserved quantities of density, momentum, and total energy, in this simulation we also evolve a passive scalar variable, $s$, which is advected with the fluid. The primary purpose of this variable is to trace where the gas in the outflow originated. Gas that is present at the start of the simulation, i.e. disk and halo gas, is initialized with a scalar value of 0. Gas that is injected in the cluster regions is given a scalar value of 1. Thus, at any later time, the fraction of mass in a cell that was originally injected via a cluster can be determined by the value of the scalar in that cell - that is, if a cell has $s = 0.5$, then half the mass in the cell was injected. Therefore, the scalar directly represents the fraction of gas in a given cell that was injected within a cluster as ``hot". Because our cluster injection accounts for unresolved interactions between the superbubble and the ISM, this value is distinct from the fraction of the mass that represents pure supernova ejecta, which varies as a function of $\beta_\mathrm{inj}$. %which would be $f_\mathrm{ej} \sim s / (10\beta_\mathrm{inj})$.

%%%%%%%%%%%%%%%%%%%%%%%%%%%%%%%%%%%%%%%%%%%%%%%%%%%%%%%%%%%%%%%%%%%%%%%

\section{Analytics} \label{sec:analytics}

Early analytic work identified hot thermalized plasma as a potential mechanism for driving gas out of galaxies at supersonic velocities \citep{Johnson71, Wolfe74}. As mentioned in Section~\ref{sec:introduction}, a particularly useful model to describe a supernova-driven wind was described by \citet{Chevalier85} (hereafter \citetalias{Chevalier85}). In brief, the \citetalias{Chevalier85} model adopted the hydrodynamic equations of mass, momentum, and energy conservation in spherical symmetry, with the addition of constant source terms for mass and energy applied within a given ``driving radius." They showed that there is an analytic steady-state solution that consists of a supersonic flow -- a fast, hot wind -- outside the driving region. Other authors have since built on this model to incorporate various additional physical effects, including gravity, radiative cooling, inflows, and non-uniform mass and energy driving regions \citep{Wang95, Efstathiou00, Silich03, Silich04, Thompson16, Bustard16}.

In this Section, we work out the expected relationships between the average physical parameters of interest (density, velocity, energy, etc.) and the mass, momentum, and energy fluxes as a function of radius for both the hot and cool phases of the outflow. We follow the approach of previous authors in applying the various continuity equations in a spherical geometry, and we allow mass and energy to transfer from one phase to another via the inclusion of source terms. In other words, we assume that these source terms are present at all radii, rather than just within the wind driving region. We begin by 
writing down the relevant equations for a single phase, which we consider the
hot phase of the wind.  We then address the effects of interactions between phases on the cool gas.

\subsection{The Hot Phase}\label{sec:hoteqs}

Unless otherwise noted, all of the variables in this subsection should be understood to have an implied `h" subscript, denoting that they refer to the hot phase of the outflow.

\textbf{Mass}. The mass continuity equation can be written 
\begin{equation}
    \partial_t(\rho) + \nabla\cdot(\rho \bm{v}) = \dot{\rho}.
\end{equation}
Here, the source term on the right hand side captures changes in mass per unit time per unit volume, and can include both mass gained in the hot phase from shocked ambient material and cool gas destruction, as well as losses from hot gas cooling out or mixing with the cool phase.
Under the assumptions of a time-steady flow and spherical symmetry, this equation becomes
\begin{equation}\label{eq:masscons}
    \frac{1}{r^{2}}d_r(r^{2}\rho v_r) = \dot{\rho}.
\end{equation}
Neither of these assumptions have to be true in an outflow - we will evaluate their validity in our simulations later. By integrating from $r = 0$ to $r = r$ over a cone, we can relate the radial profiles of density and velocity to the net mass per unit time flowing outward within the cone,
\begin{equation}\label{eqn:mass}
    \Omega r^{2} \rho v_{r} = \dot{M}_\mathrm{net},
\end{equation}
where all variables are assumed to be functions of $r$.  Here, $\Omega$ is the actual solid angle within the cone that is filled with  hot gas, i.e. allowing for a filling factor $<1$.

For clarity, we denote mass flow rates  measured at an arbitrary point in the flow as $\dot{M}_\mathrm{net}$ (and similarly rates for other quantities), to indicate that this allows for both gains and losses since the injection at the base of the flow.  
We denote rate of mass and energy injected in the cluster region as $\dot M_\mathrm{inj}$ and $\dot E_\mathrm{inj}$, respectively.

\textbf{Energy}. The energy continuity equation is
\begin{equation}
    \partial_t\left(\frac{1}{2} \rho v^2 + \frac{1}{\gamma -1} P \right) + \nabla\cdot\left(\rho \bm{v}\left(\frac{1}{2}\bm{v}^{2} + \frac{\gamma}{\gamma - 1}\frac{P}{\rho}\right)\right) = \dot{\cal{E}}_\mathrm{tot}.
\end{equation}
The energy source term on the right hand side accounts for any energy added, in addition to energy lost from cooling of the gas or mixing into the cooler phases. Assuming steady-state spherical symmetry and integrating over the cone, we get
%\begin{equation}
%    \frac{1}{r^{2}}d_r\left(r^{2}\rho v_r\left(\frac{1}{2}v_r^{2} + \frac{\gamma}{\gamma - 1}\frac{P}{\rho}\right)\right) = \dot{\cal{E}}_\mathrm{tot}.
%\end{equation}
%Integrating over the cone gives
\begin{equation}\label{eqn:energy}
    \Omega r^{2} \rho v_{r}\left(\frac{1}{2}v_r^{2} + \frac{\gamma}{\gamma - 1}\frac{P}{\rho}\right) = \dot{E}_\mathrm{net},
\end{equation}
where again, all variables are assumed to be functions of radius, and the right-hand side is the net energy per unit time flowing through the cone, consisting of the initial value in the injection region plus any gains and minus any losses.

\textbf{Scalar Mass}. As described in Section~\ref{sec:simulations}, all gas in the simulation has an associated scalar value, $s$, which is passively advected with the fluid. For material injected in a cluster, $s = 1$, while material that was originally in the disk was assigned $s = 0$. Thus, the scalar value within a given cell identifies how much of the mass in that cell was originally injected in the clusters (when $\beta_\mathrm{inj}=0.1$, this injected mass is the same as SN ejecta). We can write a continuity equation for the scalar density, $\rho_{s}$, that is identical to the density continuity equation,
\begin{equation}
    \partial_t(\rho_{s}) + \nabla\cdot(\rho_{s} \bm{v}) = \dot{\rho}_{s}.
\end{equation}
Following the same procedure as above relates the scalar mass flux to the scalar density and velocity profile:
\begin{equation}\label{eqn:color}
    \Omega r^{2} \rho_{s} v_{r} = \dot{M}_\mathrm{s,net}.
\end{equation}
The right-hand side measures the total scalar mass per unit time flowing outward in the cone, which will allow for any initial material injected, minus losses.  

An advantage of tracking this scalar mass is that it allows us to examine the relationships between total measured outflow rates at a given radius, $\dot{M}_\mathrm{net}$ and $\dot{E}_\mathrm{net}$, and the mass and energy that was injected by the clusters at the base of the flow. We note that for $r$ approaching the source region, in the idealized perfectly spherical case we would have $\dot M_\mathrm{net} = \dot M_\mathrm{s,net}=\dot M_\mathrm{inj}$.
Taking the ratio of Equation~\ref{eqn:color} to Equation~\ref{eqn:mass} gives
\begin{equation}
    \frac{\rho_{s}}{\rho} = \frac{\dot{M}_ \mathrm{s,net}}{\dot{M}_\mathrm{net}},
\end{equation}
where $\rho_s/\rho\equiv s$ is the definition of the scalar. We therefore have
\begin{equation}
\dot M_\mathrm{net} = \frac{\dot M_\mathrm{s,net}}{s},
\end{equation}
and in the case that none of the hot gas has cooled out, $\dot{M}_\mathrm{s,net}=\dot{M}_\mathrm{inj}$.   Thus, if $s<1$ in the hot medium at some large radius and there have been negligible losses to cooling, $\dot{M}_\mathrm{net}$ will exceed the initial injected value due to mixing-in  of originally-cool gas. Likewise, the ratio of Equation~\ref{eqn:energy} to Equation~\ref{eqn:color} gives
%\begin{equation}   \frac{\rho}{\rho_{c}}\left(\frac{1}{2}v_r^2 + \frac{\gamma}{\gamma - 1}\frac{P}{\rho}\right) = \frac{\dot{E}_\mathrm{net}}{\dot{M}_{c, \mathrm{in}}},\end{equation}
%or
%\begin{equation}
%    \frac{(\frac{1}{2}\bm{v}^2 + \frac{\gamma}{\gamma - 1}\frac{P}{\rho})}{c} = \frac{\dot{E}_\mathrm{net}}{\dot{M}_{c}}.
%\end{equation}
%or
\begin{equation}
    \dot{E}_\mathrm{net} = \left(\frac{1}{2}v_r^2 + \frac{\gamma}{\gamma - 1}\frac{P}{\rho}\right) \frac{\dot{M}_\mathrm{s,net}}{s}.
\end{equation}

\textbf{Momentum}. Momentum continuity states
\begin{equation}
    \partial_t(\rho \bm{v}) + \nabla\cdot(\rho \bm{v}\bm{v} + \mathrm{\textbf{I}}P) = -\rho \nabla \Phi + \dot q.
\end{equation}
While there is no (significant) momentum directly injected by the clusters, there are potentially momentum source and sink terms associated with mixing.  For example, when hot gas is mixed into the cool medium (thereafter cooling) there will be a momentum sink term (negative $\dot q$) for the hot medium. If a small enough amount of cool gas is mixed into the hot such that it does not subsequently cool, this will be a positive $\dot q$ for the hot medium.
The spherical and steady state assumptions yield
\begin{equation}\label{eq:mom1}
    \frac{1}{r^{2}}\partial_r(r^{2}\rho v_r^{2}) + \partial_r(P) = -\rho \partial_r(\Phi) + \dot q.
\end{equation}
There is no straightforward way to integrate the pressure term, so we will continue with the differential version of the momentum equation. A bit of algebra translates it into a more usable form:
%\begin{equation}
%\begin{aligned}
%\frac{r^2 \rho v_r}{r^2}\delta_r(v_r) + \frac{v_r \delta_r(r^2 \rho v_r)}{r^2} + \delta_r(P) &= -\rho \delta_r \Phi \\    \rho v_r \delta_r(v_r) + v_r \dot{\rho} + \delta_r(P) &= -\rho \delta_r \Phi,
%\end{aligned}
%\end{equation}
%or
\begin{equation}\label{eqn:momentum}
    v_r \partial_r(v_r) + \frac{\partial_r(P)}{\rho} = - \partial_r(\Phi) -\frac{\dot{\rho}}{\rho}v_r + \frac{\dot q}{\rho},
\end{equation}
from which we can assess the importance of each term in setting the velocity profile.

\textbf{Entropy}. Substituting Equation~\ref{eqn:mass} into Equation~\ref{eqn:energy} gives
\begin{equation}\label{eq:specen}
    \frac{1}{2}v_r^2 + \frac{\gamma}{\gamma - 1}\frac{P}{\rho} = \frac{\dot E_\mathrm{net}}{\dot M_\mathrm{net}}.
\end{equation}
Meanwhile, from Equation~\ref{eqn:momentum} we have
\begin{displaymath}
    \partial_r\left(\frac{1}{2}v_r^2\right) + \frac{\partial_r(P)}{\rho} =  - \partial_r(\Phi) -\frac{\dot{\rho}}{\rho}v_r + \frac{\dot q}{\rho}.
\end{displaymath}
Combining the two gives
%\begin{displaymath}
%\delta_r\left(\frac{\gamma}{\gamma-1}\frac{P}{\rho}\right) - \frac{\delta_r(P)}{\rho} = \delta_r\left(\frac{\dot{E}}{\dot{M}}\right) + \frac{\dot{\rho}}{\rho}v_r + \delta_r(\Phi).
%\end{displaymath}
%We can rewrite the left-hand side of this equation as
%\begin{displaymath}
%    \frac{1}{\rho}\left(\frac{\gamma}{\gamma-1}\right)\delta_r(P) - \frac{\gamma}{\gamma-1}\frac{P}{\rho^2}\delta_r(\rho),
%\end{displaymath}
%giving us
%\begin{displaymath}
%    \frac{1}{\gamma-1}\partial_r(P) - \frac{\gamma}{\gamma-1}\frac{P}{\rho}\delta_r(\rho) = \rho\delta_r\left(\frac{\dot{E}}{\dot{M}}\right)+\dot{\rho}v_r + \rho\delta_r(\Phi).
%\end{displaymath}
%Using the fact that 
%\begin{displaymath}
%    \delta_r(\mathrm{ln}[P \rho^{-\gamma}]) = \frac{\delta_r(P)}{P} - \frac{\gamma\delta_r(\rho)}{\rho},
%\end{displaymath} 
a differential equation describing the entropy profile:
\begin{equation}\label{eq:entropy}
    \partial_r\left(\frac{\mathrm{ln}[P\rho^{-\gamma}]}{\gamma-1}\right) = \frac{1}{P}\left[\rho\partial_r
    \left(\frac{\dot E_\mathrm{net}}{\dot M_\mathrm{net}}\right) + \rho\partial_r \Phi \\ + \frac{v_r}{r^2}\partial_r\left(\frac{\dot{M}_\mathrm{net}}{\Omega}\right)
    -\dot q \right].
\end{equation}
In general, we do not expect the gravitational term on the RHS to affect the properties for the hot phase, so it may be omitted.

\textbf{Effective $\alpha$ and $\beta$}. We can use the above relationships to determine an ``effective" $\alpha$ and $\beta$ at each radius. (Recall that $\alpha$ is the ratio of energy in the outflow to the total energy injected by supernovae, and $\beta$ is the ratio of mass flux in the outflow to star formation rate.) 

We can define an effective mass loading in the wind at any radius by
\begin{equation}
    \beta_\mathrm{eff} \equiv \frac{\dot{M}_\mathrm{net}}{\dot{M}_\mathrm{SFR}} 
    =\beta_\mathrm{inj} \frac{\dot{M}_\mathrm{net}}{\dot{M}_\mathrm{inj}}
   =\frac{\beta_\mathrm{inj}}{s}\frac{\dot{M}_\mathrm{s,net}}{\dot{M}_\mathrm{inj}}.
\label{eqn:eff_beta}    
\end{equation}
Similarly,
\begin{equation}
    \alpha_\mathrm{eff} \equiv \frac{\dot{E}_\mathrm{net}}{E_\mathrm{SN}(\dot{M}_\mathrm{SFR}/m_{*})}=\alpha_\mathrm{inj} \frac{\dot E_\mathrm{net}}{\dot E_\mathrm{inj}},
\label{eqn:eff_alpha1}
\end{equation}
where $E_\mathrm{SN}$ is the energy injected per supernova, $m_{*}$ is the mass of stars formed per supernova, and $\dot{M}_\mathrm{SFR}$ is the total star formation rate.
We can use Equation~\ref{eq:specen} to rewrite this as
\begin{equation}
    \alpha_\mathrm{eff} = \left(\frac{1}{2}v_r^2 + \frac{\gamma}{\gamma - 1}\frac{P}{\rho}\right)\frac{m_{*}}{E_\mathrm{SN}}\frac{\dot M_\mathrm{s,net}}{\dot M_\mathrm{inj}} \frac{\beta_\mathrm{inj}}{s}.
\label{eqn:eff_alpha}
\end{equation}
%or alternatively 
%\begin{equation}
%    \alpha_\mathrm{eff} =    \frac{\alpha_\mathrm{in}}{c}   \frac{\dot M_\mathrm{c,net}}{\dot M_\mathrm{in}}     \frac{\dot E_\mathrm{net}/\dot E_\mathrm{in}}{\dot M_\mathrm{net}/\dot M_\mathrm{in}}.
%\end{equation}
Hence, the scalar variable allows us to distinguish between the injected mass and energy rates set by our cluster prescription, and the effective mass and energy loading actually measured in the simulation, at any point in the wind.

In the absence of cooling, $\dot E_\mathrm{net}=\dot E_\mathrm{inj}$ and there is no reduction of total scalar mass in the hot medium with distance, so that $\dot  M_\mathrm{s,net} = \dot  M_\mathrm{s,inj}=\dot M_\mathrm{inj}$. In this case, we have  
$\alpha_\mathrm{eff}=\alpha_\mathrm{inj}$ and
$\beta_\mathrm{eff} = \beta_\mathrm{inj}/s$. That is, the effective  mass loading can be different from the injected value due to  mixing of previously-cool gas into  the hot medium. 

\subsection{The Cool Phase}

Completely analogous equations to those of Section \ref{sec:hoteqs} could be written for the cool gas.  However, beyond the injection region, we know that for mass, scalar mass, and momentum, there are no ``exogenous'' sources or sinks.  Thus, any losses from the hot must be gains for the cool, and vice versa.  That is, $\dot \rho_\mathrm{c} = -\dot \rho_\mathrm{h}$, $\dot \rho_{s,\mathrm{c}}  = - \dot \rho_{s,\mathrm{h}}$, and $\dot q_\mathrm{c} = -\dot q_\mathrm{h}$. For the energy, however, total energy summed over phases is not conserved due to radiation.

In order to accelerate the cool gas, momentum must be transferred to it from the hot phase. We assume that the cool gas is supersonic, and therefore we can neglect the pressure term in the ``cool'' version of Equation \ref{eq:mom1}. The continuity equation for momentum then states 
\begin{displaymath}
    \left[\frac{1}{r^2}\partial_r(r^2\rho v_r^2) + \rho\partial_r(\Phi)\right]_\mathrm{c} = - \left[\frac{1}{r^2}\partial_r(r^2 \rho v_r^2) + \partial_r(P)\right]_\mathrm{h},
\end{displaymath}
where the source term on the right-hand side is the rate of momentum transferred per unit volume from the hot gas (subscript h) to the cool gas (subscript c). This is $\dot q_\mathrm{c} = -\dot q_\mathrm{h}$, and we have neglected the gravitational potential term for the hot medium. We can rewrite the right-hand side of this equation as
%\begin{displaymath}
%    -\left[\frac{v_r}{r^2}\delta_r(r^2 \rho v_r) + \rho v_r \delta_r(v_r) + \delta_r(P)\right]_\mathrm{h}
%\end{displaymath}
%\begin{displaymath}
%   = -\left[v_r \dot{\rho} +  \rho v_r \delta_r(v_r) + \delta_r(P)\right]_\mathrm{h}
%\end{displaymath}
\begin{displaymath}
   = -\left[v_r \frac{\dot{\rho_s}}{s} +  \rho v_r \partial_r(v_r) + \partial_r(P)\right]_\mathrm{h},
\end{displaymath}
where we  have  used $\dot \rho_h =  \dot \rho_{s,\mathrm{h}}/s_\mathrm{h}$.
Conservation of the scalar variable combined with mass conservation for the cool (cf. Equation \ref{eq:masscons}) yields
\begin{displaymath}
-\dot{\rho}_{s,\mathrm{h}}= \dot{\rho}_{s,\mathrm{c}} =   \left[\frac{1}{r^2}\partial_r(r^2 \rho_s v_r)\right]_\mathrm{c}. 
\end{displaymath}
Thus, we can rewrite the momentum equation as
\begin{equation}
\begin{split}
    \partial_r(r^2 \rho v_r^2)_\mathrm{c} = - r^2 \rho_\mathrm{c}\partial_r(\Phi) + \frac{v_{r,\mathrm{h}}}{s_\mathrm{h}}\partial_r(r^2 \rho_s v_r)_\mathrm{c} \\ - r^2\left[\rho v_r \partial_r(v_r) + \partial_r(P)\right]_\mathrm{h}.
\end{split}
\label{eqn:warm_momentum}
\end{equation}
The  left-hand side is the rate of increase of momentum flux in the cool gas as a function of $r$, while the first term on the right-hand side accounts for deceleration of the cool gas due to gravity. The second term on the right-hand side describes the rate of increase of the cool momentum flux transferred from the hot medium, at a rate proportional to the rate of increase of scalar in the cool gas resulting from mixing in of the hot gas. The last term on the right-hand side is the thermal pressure work from the hot on the cool.  The second-to-last term on the right-hand side describes the deceleration of the hot gas; this is term that represents the ram pressure work of the hot on the cool. If the hot gas is a supersonic wind, its deceleration and pressure gradients will be small. 

If we assume that the mixing is the dominant source term  for the cool gas momentum flux, and that the hot wind velocity and scalar have reached constant values, Equation \ref{eqn:warm_momentum} gives a linear relationship between the velocity of the cool gas and its scalar value:
\begin{equation}
    \rho_\mathrm{c} v_{r, \mathrm{c}}^2 = \frac{v_{r, \mathrm{h}}}{s_\mathrm{h}}\rho_{s,\mathrm{c}} v_{r,\mathrm{c}},
\end{equation}
or
\begin{equation}\label{eq:warmvelpred}
    v_{r, \mathrm{c}} = \frac{v_{r, \mathrm{h}}}{s_\mathrm{h}} s_\mathrm{c},
\end{equation}
which is normalized by the velocity of the hot gas.

%%%%%%%%%%%%%%%%%%%%%%%%%%%%%%%%%%%%%%%%%%%%%%%%%%%%%%%%%%%%%%%%%%%%%%%

\section{Results} \label{sec:results}

We first present a qualitative overview of the simulation, focusing on snapshots at a few characteristic times. Following the overview, we expand on the radial dependence of density, velocity, pressure, and temperature in Section~\ref{sec:profiles}; as well as the radial mass, momentum, and energy fluxes split by gas phase in Section~\ref{sec:fluxes}. We then show phase diagrams and their relationship to the gas velocity in Section~\ref{sec:histograms}, and finish with a discussion of convergence in our model (Section~\ref{sec:convergence}).

\subsection{Simulation Overview}\label{sec:overview}

\begin{figure}[b]
    \centering
    \includegraphics[width=
    \linewidth]{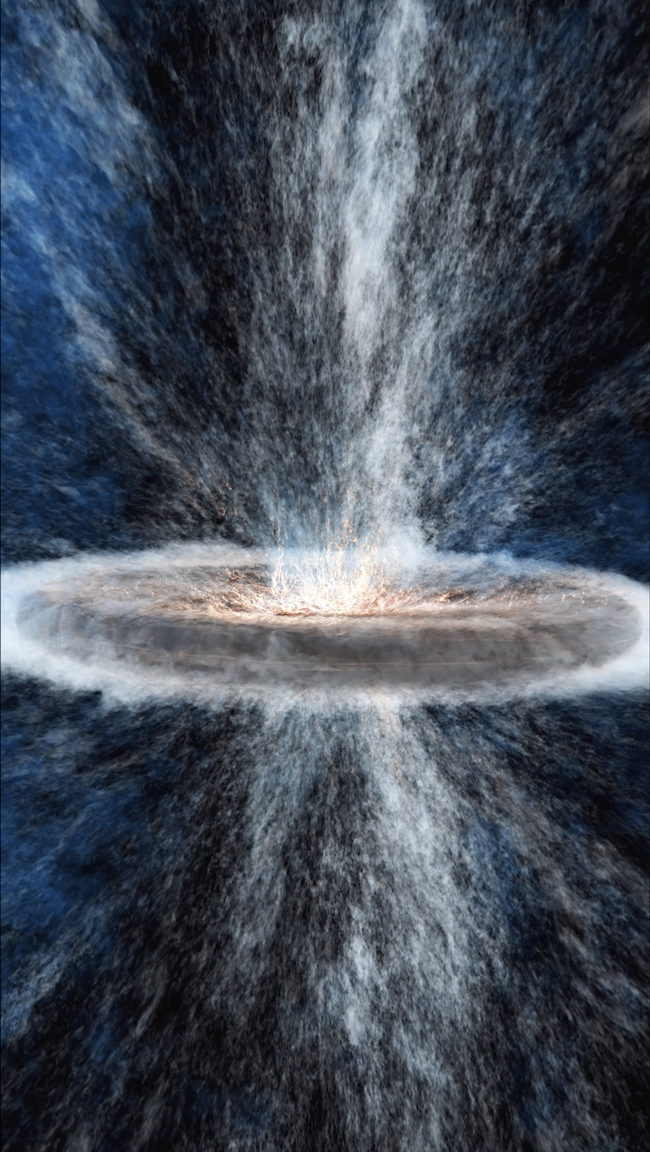}
    \caption{Rendering of the density field at 35~Myr, highlighting the disk, high density clouds being driven out at the center, and lower density, more diffuse clouds at larger radii. The highest density gas is peach, while lower density more diffuse gas is white, and the lowest densities are blue/black. Figure made using the NVIDIA IndeX software.}
    \label{fig:index_fig}
\end{figure}

\begin{figure*}
    \centering
    \includegraphics[width=0.4\linewidth]{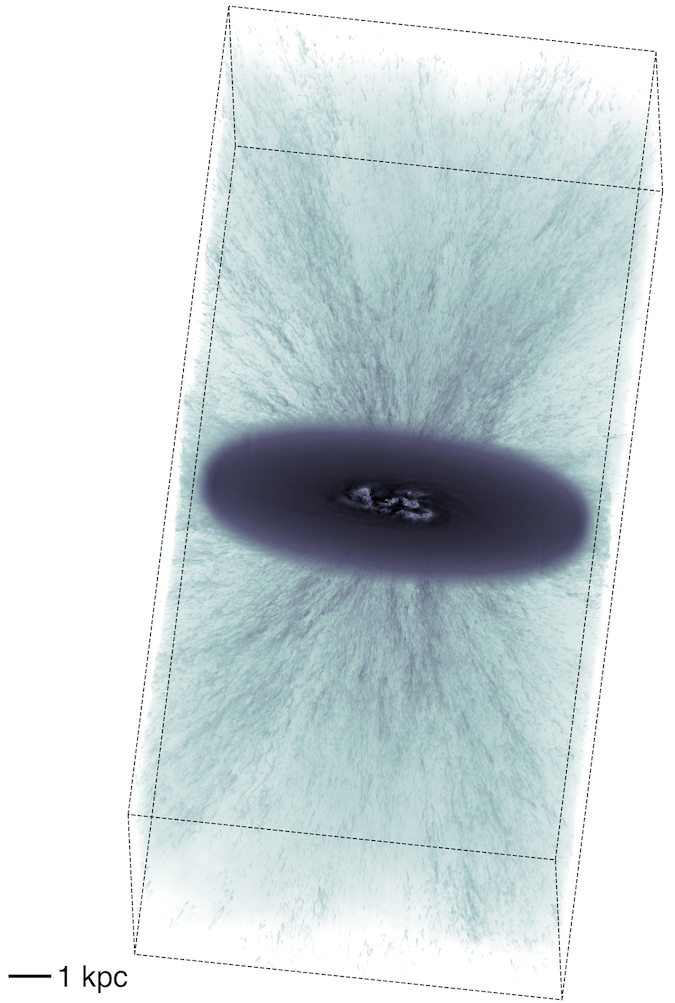}
   \vspace{1.5cm}
    \includegraphics[width=0.4\linewidth]{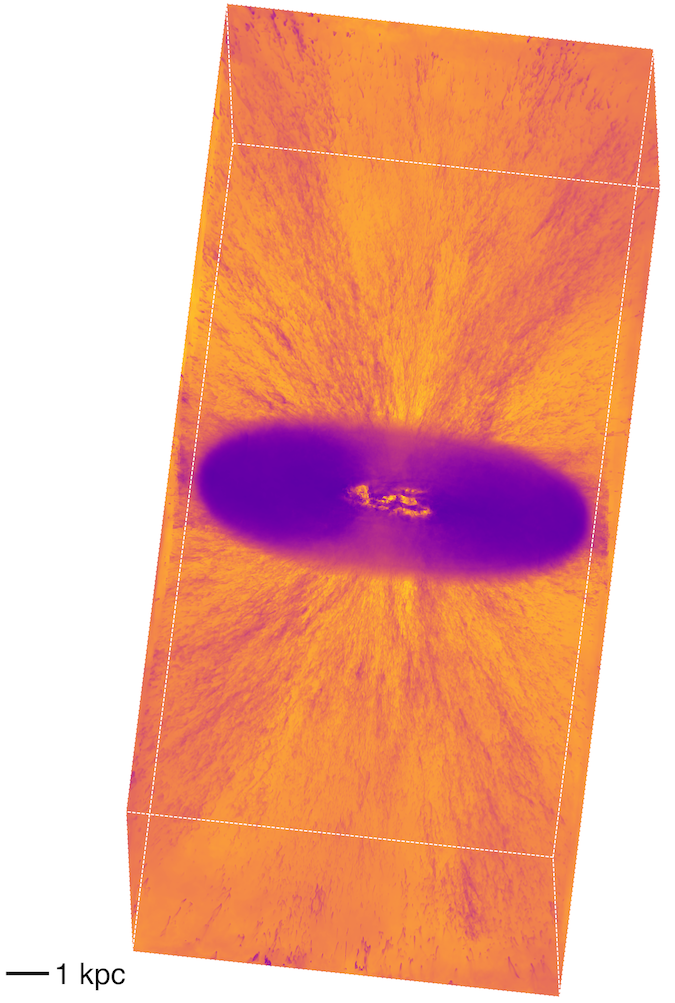}    
    \caption{Off-axis density and density-weighted temperature projections of the full simulation volume at 35 Myr.}
    \label{fig:projections}
\end{figure*}

As with previous CGOLS simulations, we endeavor to make the most of this (expensive) simulation by running in two different states. We begin the cluster feedback at 5~Myr in the high star formation rate state, with $\dot{M}_\mathrm{SFR} = 20\,\mathrm{M}_\odot\,\mathrm{yr}^{-1}$, equivalent to turning on 2 clusters every million years for 30~Myr. After that point, we turn on clusters at a rate corresponding to $\dot{M}_\mathrm{SFR} = 5\,\mathrm{M}_\odot\,\mathrm{yr}^{-1}$, our low star formation rate state. These values were chosen to match our previous simulations. In the following, we primarily present results for two characteristic times - at 35~Myr, when the SFR has been high for its maximum time, and at 65~Myr, when it has been low for the same length of time. 

Figure~\ref{fig:index_fig} shows a rendering of the density field at 35~Myr that highlights many general features of the outflow. At small radii, high density clouds of disk gas are being driven out by the central clusters. At larger radii, the outflowing material is more diffuse, but there are still high density filaments permeating the volume filling low density phase. These high density filaments are clearly associated with the cool gas in the outflow, as can be seen in Figure~\ref{fig:projections}, which shows off-axis density and temperature projections of the full volume at 35~Myr. Movies showing the density and temperature projections for the full time evolution can be found online\footnote{Movies showing time evolution of all the CGOLS simulations are located at http://evaneschneider.org/simulation-gallery}.

\begin{figure*}
    \centering
    \includegraphics[width=0.4\linewidth]{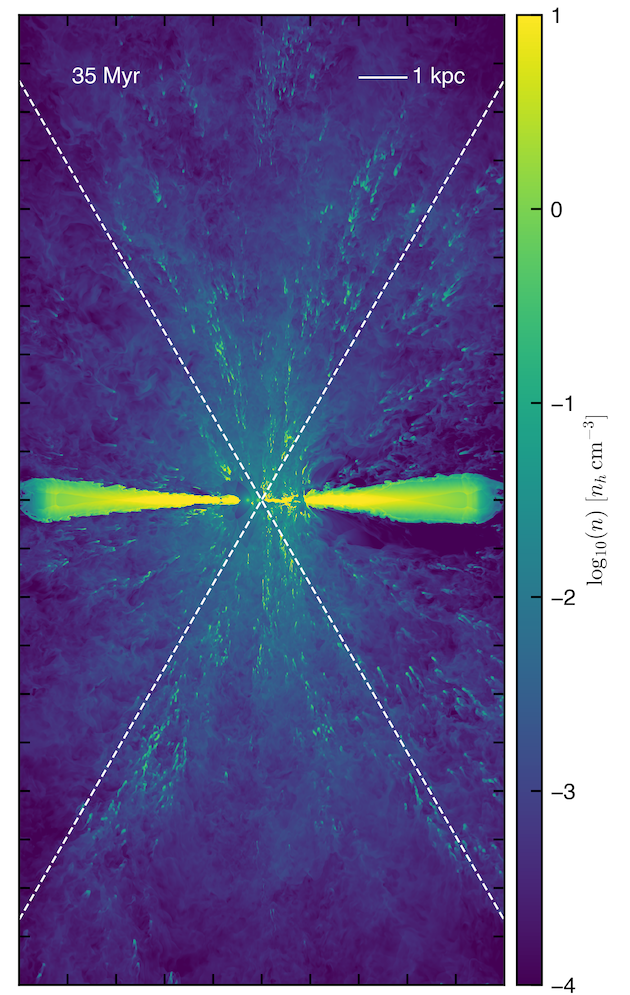}
    \includegraphics[width=0.4\linewidth]{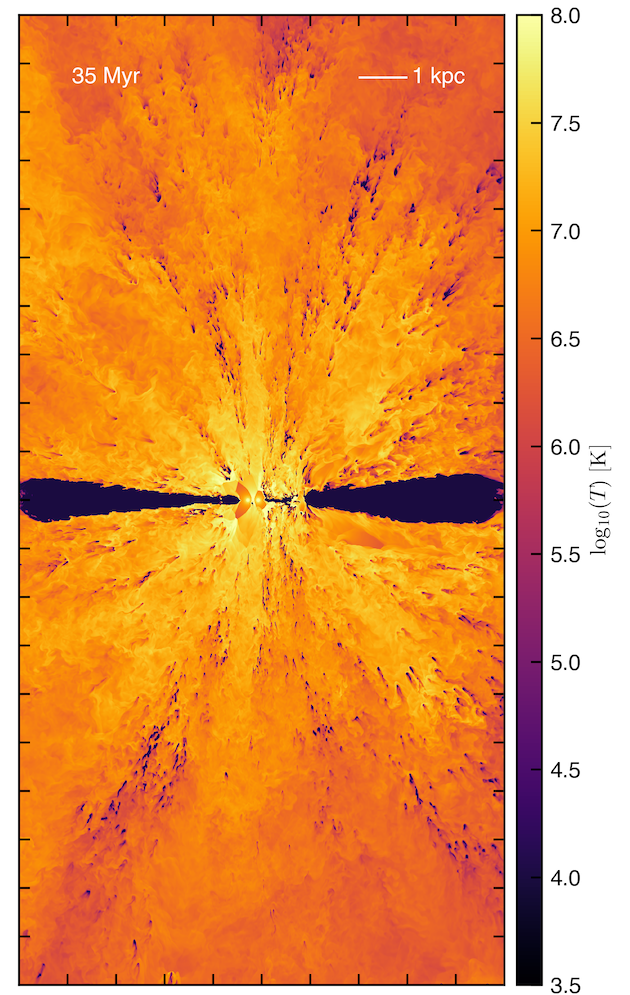}
    \includegraphics[width=0.4\linewidth]{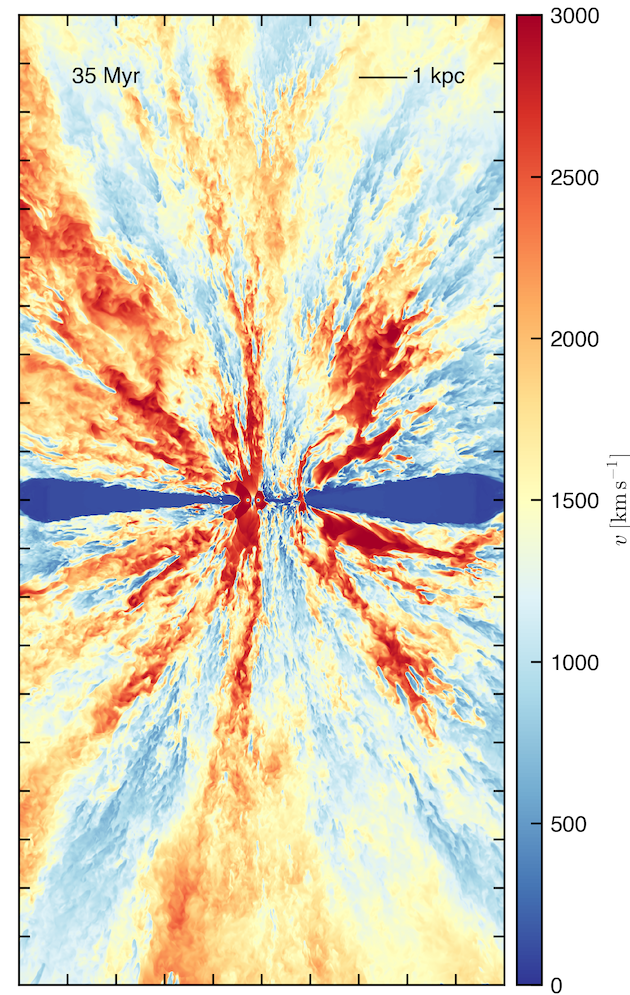}
    \includegraphics[width=0.4\linewidth]{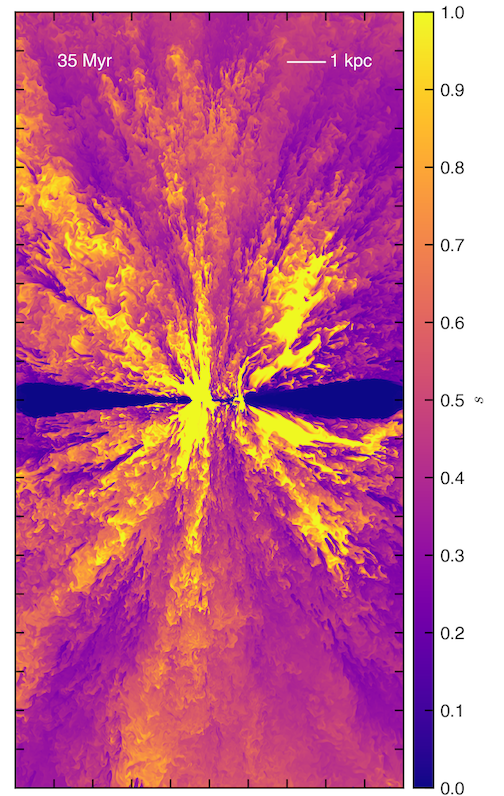}
    \caption{Number density, temperature, velocity, and scalar slices through the $y = 0$ plane at 35 Myr.}
    \label{fig:slices_35}
\end{figure*}

\begin{figure*}
    \centering
    \includegraphics[width=0.4\linewidth]{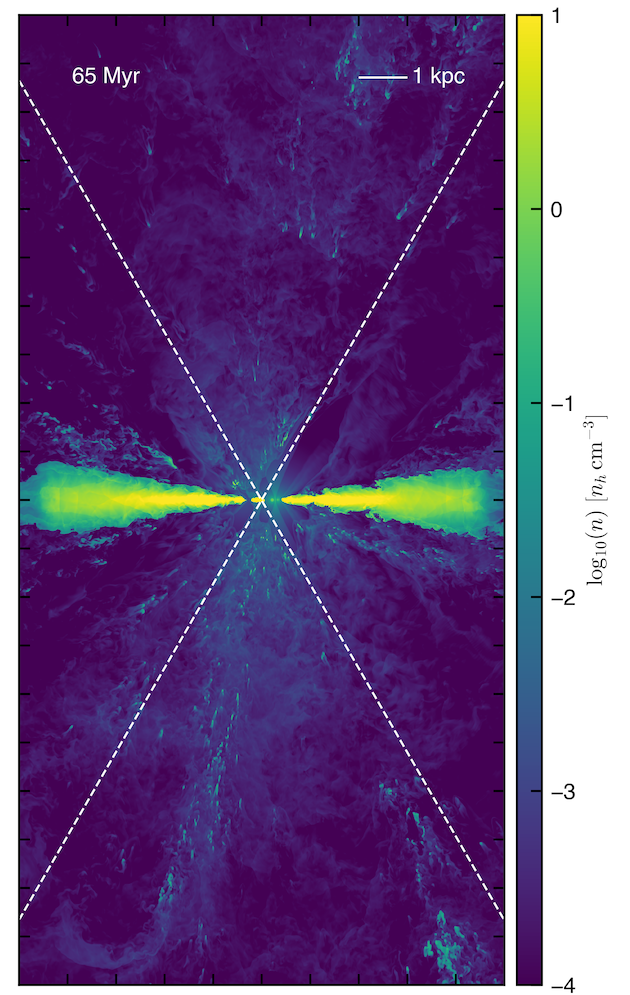}
    \includegraphics[width=0.4\linewidth]{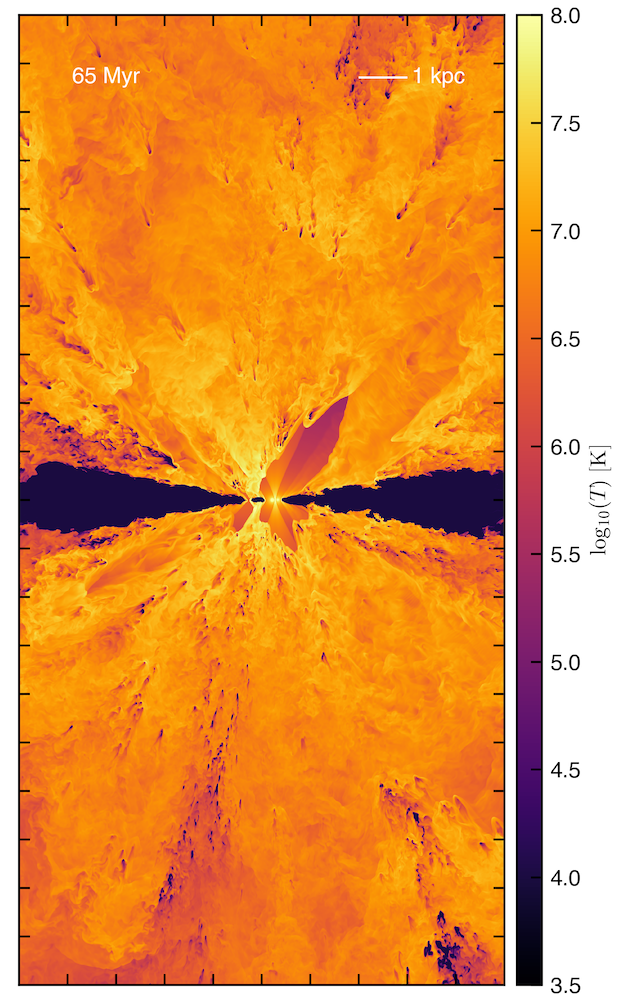}
    \includegraphics[width=0.4\linewidth]{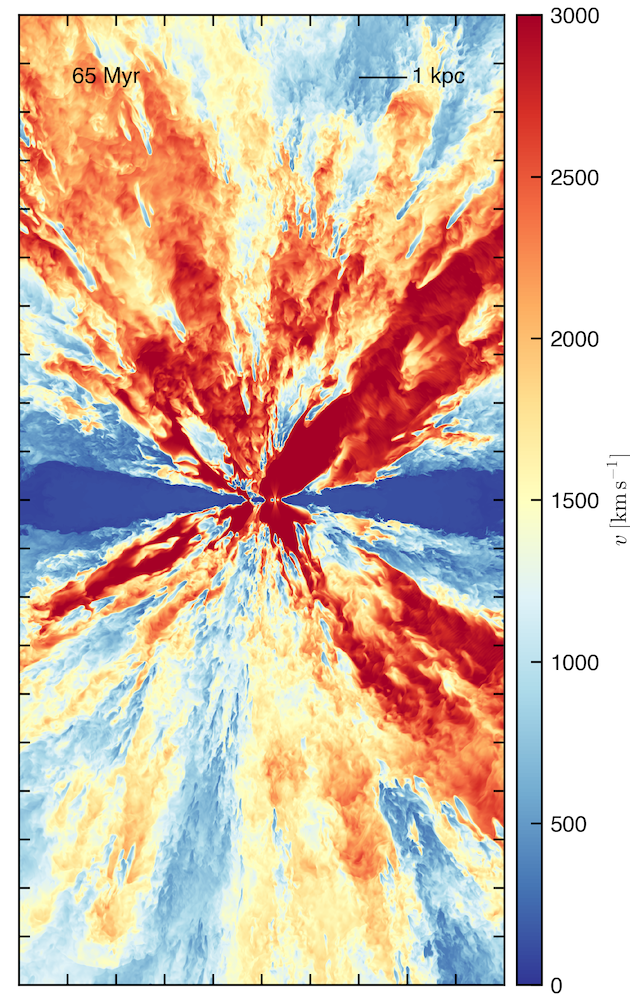}
    \includegraphics[width=0.4\linewidth]{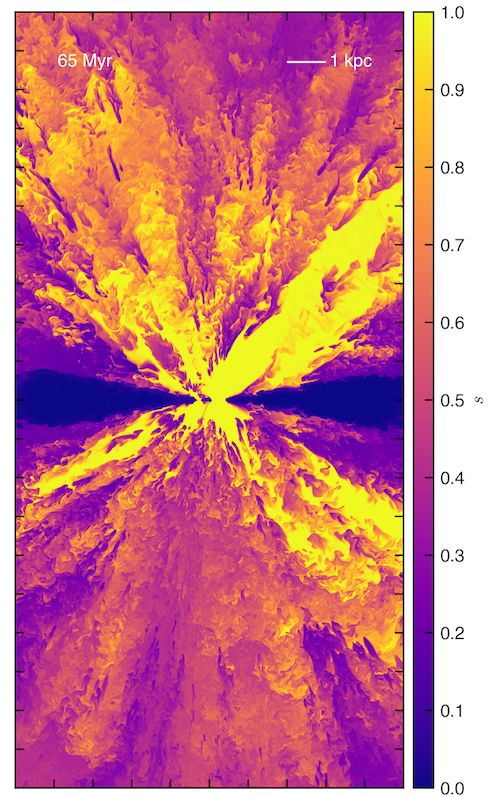}
    \caption{Number density, temperature, velocity, and scalar slices at 65 Myr.}
    \label{fig:slices_65}
\end{figure*}

In addition to the projections, slices through the box reveal interesting relationships between the density, velocity, temperature, and scalar values of gas in the outflow. In Figure~\ref{fig:slices_35} we show density, temperature, velocity, and scalar slices along the $x-z$ plane during the high SFR state, while Figure~\ref{fig:slices_65} shows the same slices during the low SFR state. A few salient features of the outflow can be determined immediately upon inspection of these slices. First, at both times, the outflow is multiphase, characterized by a volume-filling hot phase at $T > 10^6\,\mathrm{K}$ punctuated by small, dense clouds of cool gas at $T\sim10^4\,\mathrm{K}$. There is a clear correlation between the gas density, velocity, and temperature, with the lower density channels corresponding to the hottest, fastest-moving gas. The outflow features are roughly biconical, though determining an opening angle is not straightforward, and would likely depend on which snapshot was being examined. Regardless, the opening angle appears to be large, and the outflow does not show evidence of any sort of fountain, with cool gas raining back down onto the disk at larger radii. We note that this may be a result of our choice to position the clusters within the central $R = 1\,\mathrm{kpc}$ of the disk, and that our limited horizontal volume prevents us from assessing whether a fountain flow would arise at large angles at larger radii.

In contrast with our previous simulations, at no point in this simulation does the volume-filling hot phase become mass-loaded enough to undergo significant radiative cooling. While the simulations in \citetalias{Schneider18b} employed an arbitrary $\beta_\mathrm{inj} = 0.4$ for the hot phase, in this simulation we set the mass-loading for the injected material to a physically-motivated value determined by massive star winds, SN ejecta, and cluster breakout. This average $\beta$ is close to 0.1 most of the time, and mixing with the disk gas at small radii is not efficient enough to increase the density of the majority of the hot gas enough for it to cool. This does not mean none of the hot gas cools, however. As we will demonstrate later, there is strong evidence that hot gas that does interact with cool clouds creates a mixed phase with intermediate temperatures, which can cool under the right circumstances.

Comparing Figure~\ref{fig:slices_35} to Figure~\ref{fig:slices_65}, we see that overall the simulation during the higher SFR state is characterized by a higher density, lower velocity outflow, and the biconical outflow region is filled with more clouds of cool gas. At late times when the SFR is lower, there is not enough interaction between the clusters and the disk gas to contribute significant mass-loading to the outflow, as we will show directly when we examine the fluxes in Section~\ref{sec:fluxes}. This lower mass-loading leads to a lower-density hot phase, and fewer cool clouds in the outflow.

\subsection{Radial Profiles}\label{sec:profiles}

\begin{figure*}
    \centering
    \includegraphics[width=\linewidth]{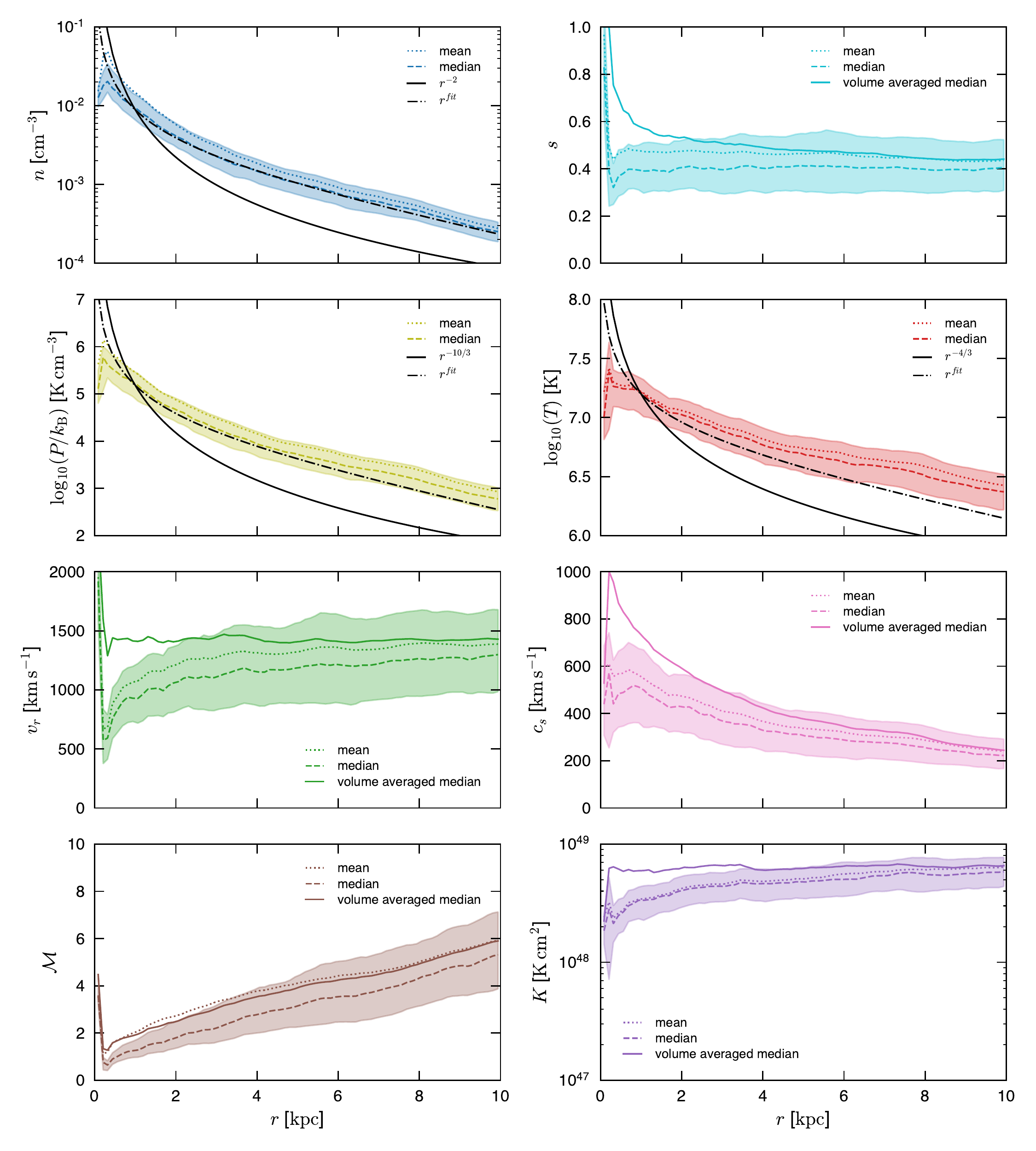}
    \caption{Density-weighted average radial profiles for the hot phase ($T > 5\times10^5\,\mathrm{K}$) at 35 Myr within a biconical region with a half opening angle of $30\degree$. From top left to bottom right, the profiles show number density, scalar, pressure, temperature, radial velocity, sound speed, Mach number, and entropy. In addition to the mean and median, upper and lower quartiles of the distributions are shown. For profiles where they differ significantly, median values for the volume-averaged quantities are also displayed. Solid black lines in the density, pressure, and temperature panels show the expected radial profiles assuming adiabatic expansion, normalized to the average values from the simulation at 1 kpc. Dashed black lines show alternative profiles, accommodating additional mass transfer into the hot phase at all radii.}
    \label{fig:profiles_hot}
\end{figure*}

\begin{figure*}
    \centering
    \includegraphics[width=\linewidth]{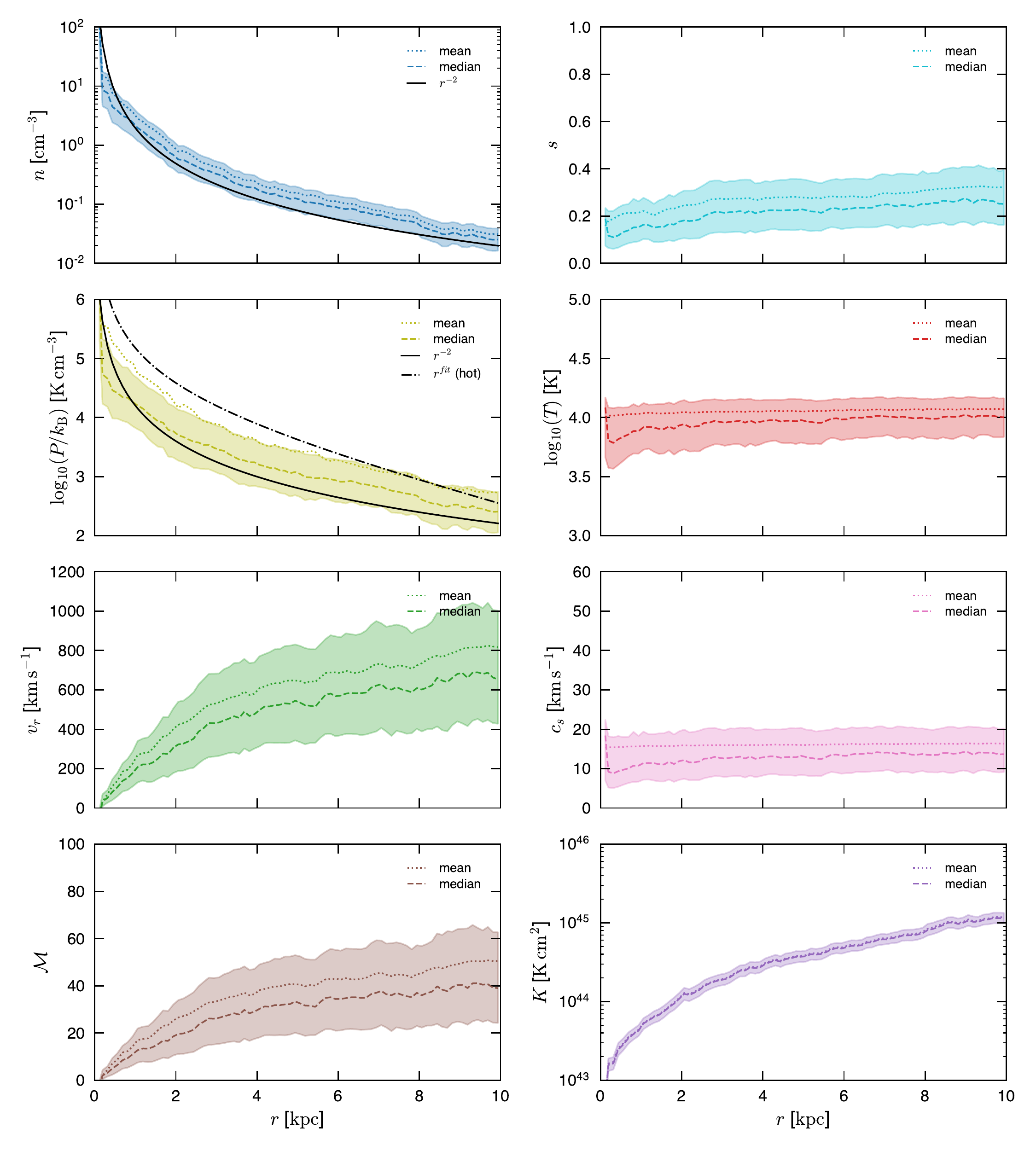}
    \caption{Density-weighted average radial profiles for the cool phase ($T < 2\times10^4\,\mathrm{K}$) within a $30\degree$ cone at 35 Myr. From top left to bottom right, the profiles show number density, scalar, pressure, temperature, radial velocity, sound speed, mach number, and entropy. Black solid lines show the expected density and pressure scalings for isothermal radial expansion, while the black dashed line in the pressure panel shows the best-fit scaling from Figure~\ref{fig:profiles_hot}.}
    \label{fig:profiles_cool}
\end{figure*}

Throughout the remainder of the results, we will examine properties of the gas split by phase. We define three phases of interest: hot gas ($T > 5\times10^5\,\mathrm{K}$), intermediate temperature gas ($2\times10^4\,\mathrm{K} < T < 5\times10^5\,\mathrm{K}$), and cool gas ($T < 2\times10^4\,\mathrm{K}$). Most of our analysis focuses on the hot and cool phases, as they are the longest-lived and represent the majority of the mass in the outflow (see Section~\ref{sec:histograms}). 

In Figures~\ref{fig:profiles_hot} and \ref{fig:profiles_cool} we show radial profiles for the hot and cool phases, respectively, for a number of physical parameters of interest: the number density $n$, passive scalar $s$, pressure $P$, temperature $T$, radial velocity $v_r$, sound speed $c_s$, Mach number $\mathcal{M}$, and entropy $K \propto P \rho^{-\gamma}$, where $\gamma$ is the adiabatic index of the gas, taken to be $5/3$ through this work. Note that in Figures~\ref{fig:profiles_hot} and \ref{fig:profiles_cool}, we plot the entropy as $K \propto T n^{-\frac{2}{3}}$. Each profile is measured within a cone with half-opening angle $\Omega = 30\degree$ above and below the disk, as shown with dashed white lines in the density panel of Figures~\ref{fig:slices_35} and \ref{fig:slices_65}. Within the cone, we expect the properties of the outflow to be approximately spherically symmetric once we reach a radius equal to our cluster seeding radius within the disk, $R = 1\,\mathrm{kpc}$.

In each panel, we show the median and mean values, as well as the 1st and 3rd quartiles, to give an idea of the spread in the gas properties. We use density-weighted values for all quantities, unless otherwise noted. For example, the velocity average is calculated as $\langle v_{r} \rangle = \sum (v_r n) / (\langle n \rangle N_\mathrm{cells})$, with $\langle n \rangle = \sum (n) / N_\mathrm{cells}$ for all cells within the cone that meet the relevant temperature criterion. The choice of density-weighted versus volume-weighted averages has no effect on the cool gas profiles, but there are some differences for the hot phase. Thus, we also include the volume-averaged median quantities on relevant panels in Figure~\ref{fig:profiles_hot}. However, we note that the largest differences are at small radii, where we do not interpret our results as complete, because we are still within the gain region. The biggest differences between the mass-weighted and volume-weighted profiles arise in the scalar variable, which tends to be higher in the hot gas if we use a volume-weighted average for the hot phase, and the velocities, which are also higher on average. We further explore the relationship between the scalar and velocity in Section~\ref{sec:acceleration}.

\textit{The hot phase}. On average, the profiles for the hot phase follow relationships that are close to, but not exactly, what one would expect given adiabatic expansion of the gas. For example, \cite{Chevalier85} calculated the radial solution for a hot outflow given mass and energy injection within a spherical region. In that model, density decreases as $r^{-2}$, while velocity quickly asymptotes to a value set by $v_\mathrm{term} = \sqrt{2 \dot{E}_\mathrm{inj} / \dot{M}_\mathrm{inj}}$, as would be predicted by Equation \ref{eq:specen} (assuming mass and energy are conserved after injection). With an adiabatic index $\gamma = 5/3$, adiabatic cooling implies that pressure decreases as $r^{-10/3}$ (as Equation \ref{eq:entropy} implies when source terms are zero), and thus temperature decreases as $r^{-4/3}$. We have plotted these relationships in black lines on the panels in Figure~\ref{fig:profiles_hot} to demonstrate their deviations, with the adiabatic profiles normalized to the measured values at 1 kpc. As the first panel shows, the density is falling off with radius, but at a shallower rate than $r^{-2}$. This reflects the fact that mass is being added to the hot phase as a function of radius, as we will show in Section~\ref{sec:fluxes}. Similarly, the pressure and temperature are decreasing with radius, but at a slower rate than is implied by pure adiabatic expansion.

If we account for the mass addition to the hot phase by allowing a shallower density profile, we can accommodate changes in the pressure and temperature profiles, as well. For example, we can scale the radial dependence of the density profile such that it is an arbitrary function of $r$ that provides a good fit to the density profile. In the case of the plots in Figure~\ref{fig:profiles_hot}, we use $n \propto r^f$, with $f = -0.05 r - 1.08$, and $r$ measured in kpc. Then $P \propto r^{f \gamma}$, and $T \propto r^{f (\gamma - 1)}$. We have plotted these additional relationships as black dashed lines in Figure~\ref{fig:profiles_hot}. While the slope of the density profile is now an arbitrary function, the slopes of the pressure and temperature profiles are set by the adiabatic physics, and provide a much better fit to the data than the pure expansion wind model. In this framework the hot phase of the wind can be understood entirely as adiabatic expansion with a mass source that depends on mixing in gas from cooler phases.

Although these new scalings do provide a much better fit to the density, pressure, and temperature profiles, there is still clearly a deficit in the pressure and temperature. The entropy profiles displayed in the final panel of Figure~\ref{fig:profiles_hot} provide an explanation. While the volume-averaged entropy profile for the hot gas is quite flat, the density-weighted median entropy in the hot phase is an increasing function of radius. This increasing slope reflects the fact that low entropy gas from the cool phase is being mixed into the hot gas, with the highest-density hot gas having experienced the most mixing. As the cool gas is added to the hot, its entropy rises, as would be expected in the case where the cool gas experiences a shock. This rising entropy profile then violates the assumption that the relationship between $n$, $P$, and $T$ is adiabatic, and in particular will result in a flatter temperature slope than predicted by adiabatic physics.

The expectation for the terminal value of the velocity is set by our choice of input $\dot{E}$ and $\dot{M}$. If we use our values of $\alpha_\mathrm{inj}$ and $\beta_\mathrm{inj}$, we find that $v_\mathrm{term} = \sqrt{2 \dot{E} / \dot{M}} \approx 3000\,\mathrm{km}\,\mathrm{s}^{-1}$, about a factor of 2 too high. However, if we measure the effective $\alpha$ and $\beta$ in the wind at $R = 1\,\mathrm{kpc}$, we find that $\alpha_\mathrm{eff} = 0.5$, while $\beta_\mathrm{eff} = 0.2$ (see Section~\ref{sec:loading}), which leads to a terminal velocity of $v_\mathrm{term} \approx 1500\,\mathrm{km}\,\mathrm{s}^{-1}$, approximately the measured velocity in the volume-averaged hot phase. The hot gas velocity as measured by the density-weighted median is slightly lower, which can naturally be explained by higher energy losses associated with higher density gas (resulting in a lower $\alpha_\mathrm{eff}$).

Another direct indication of the mass transfer between phases is the radial profile of the scalar variable in both Figures~\ref{fig:profiles_hot} and \ref{fig:profiles_cool}. By a radius of 1 kpc, the median density-weighted scalar value of the hot phase has already decreased from 1 to a value of 0.4 - indicating significant mixing from the cool phase. Similarly, the median scalar value in the cool phase has risen from 0 to 0.15. At first look, the relatively flat value of the density-averaged scalar in the hot phase combined with the rising scalar value in the cool phase might seem to be discrepant. However, the volume-averaged scalar in the hot phase tells the full story. Here, we see a rapid decrease down to $s = 0.5$ at $R < 2\,\mathrm{kpc}$. At larger radii, $s$ continues to decrease, although at a slower rate. This is consistent with rapid mixing between the hot and cool gas as the clusters initially accelerate cool material out of the disk, followed by more gradual mixing at larger radii as the volume density of cool gas goes down. The outflow rates presented in Section~\ref{sec:fluxes} bear out this explanation. However, we caution that this interpretation is dependent on the assumption that the outflow is steady-state. While we have purposely selected a time for this analysis where that appears to be the case, it is certainly not true for the entire simulation.

\textit{The cool phase}. The profiles for the cool phase can be interpreted as a radially expanding isothermal gas. Density falls off approximately as $n \propto r^{-2}$, and the temperature is a constant value set by our temperature floor. (In reality, we expect photoionization from the local ionizing sources within the galaxy to play this role at small radii, and from the cosmic UV background to provide heating of the cool medium at larger radii.) Pressure also falls off approximately as $P \propto r^{-2}$. However, the pressure in the cool phase is somewhat lower than that in the hot phase - by $\sim 0.5$~dex at 1 kpc, and with larger differences at smaller radii. We highlight this by plotting the fit to the hot phase pressure from Figure~\ref{fig:profiles_hot} on the pressure panel in Figure~\ref{fig:profiles_cool}. 

Physically, the lower pressure of the cool phase can be understood if the cooling time for intermediate temperature gas in individual cool clouds is shorter than the time it takes for them to equilibrate to the background wind pressure via pressure waves or shocks. In this scenario, cool gas in the simulation continuously interacts with the hot phase via mixing or shocks. If the interaction heats the cool gas to temperatures at the low end of the intermediate temperature range, the relatively high density leads to very fast cooling. Because the interactions between the hot and cool phase are frequent, there is insufficient time for the cool gas to equilibrate to the background hot pressure before experiencing a subsequent interaction. In particular, given the average values of the profiles at 1~kpc, the sound crossing time for cool gas at our resolution limit $\Delta x = 5\,\mathrm{pc}$ is $t_\mathrm{sc} \approx 5\times10^5\,\mathrm{yr}$, which is an order of magnitude longer than the cooling time of gas at a density of $n\sim\,1\,\mathrm{cm}^{-3}$ (see Section~\ref{sec:histograms}). The relevant timescale may instead be the slightly shorter cloud crushing time, $t_\mathrm{cc} = (n_\mathrm{c} / n_\mathrm{h})^{1/2} (\Delta x / v_\mathrm{h}) \approx 10^5\,\mathrm{yr}$ \citep{Klein94}, but this is still longer than the cooling time. In either case, the cool gas would then experience compression from the hot phase, as it is out of equilibrium. The fact that the slopes of both the density and pressure for the cool phase are slightly shallower than $r^{-2}$ could be explained by this compression. We note that it is plausible that with higher resolution, the cool clouds may further ``shatter" \citep{McCourt18}, leading to faster equilibration with the pressure of the hot phase. This explanation is consistent with our previous work investigating cool clouds at much higher resolution \citep{Schneider17}.

An important implication of this work is shown in the velocity panel of Figure~\ref{fig:profiles_cool}. In this panel, we see that the velocities of the cool gas are a rising function of radius, with maximum velocities reaching $1000\,\mathrm{km}\,\mathrm{s}^{-1}$. In fact, almost all cool gas at large radius is moving with velocities of at least $400\,\mathrm{km}\,\mathrm{s}^{-1}$, which is approximately the escape velocity of the system. The scalar value of the cool gas is also increasing with radius; we will explore the origin of the cool gas acceleration in Section~\ref{sec:acceleration}.

\subsection{Outflow Rates}\label{sec:fluxes}

As  discussed in Section~\ref{sec:analytics}, the radial profiles of  the quantities  shown in Section~\ref{sec:profiles} are related to total mass and energy outflow rates. We have measured these outflow rates in the simulation within the same cone used to calculate the radial profiles. Each flux is averaged within a radial bin of width $\Delta r = 0.125\,\mathrm{kpc}$. So to calculate the mass flux, for example, we sum the total $\dot{M}$ in a given phase within the cone between $r$ and $r + \Delta r$, and then divide by the bin width: $\dot{M} = \sum (M v_{r}) / \Delta r$, where the sum is taken over individual cells. Our cone opening angle corresponds to a solid angle of $\Omega = 2\times2\pi[1 - \mathrm{cos}(30\degree)] \approx 1.68$, or just over $\nicefrac{1}{8}$ of the full spherical area. If the outflow were perfectly spherical, then we might expect these fluxes to be $\nicefrac{1}{8}$ of the total. However, when we measure over the full $4\pi$, we find fluxes that are only a factor of $\sim 3$ larger. Hence, we conclude that the outflow is preferentially biconical, with a modest opening angle.

\begin{figure}
    \centering
    \includegraphics[width=\linewidth]{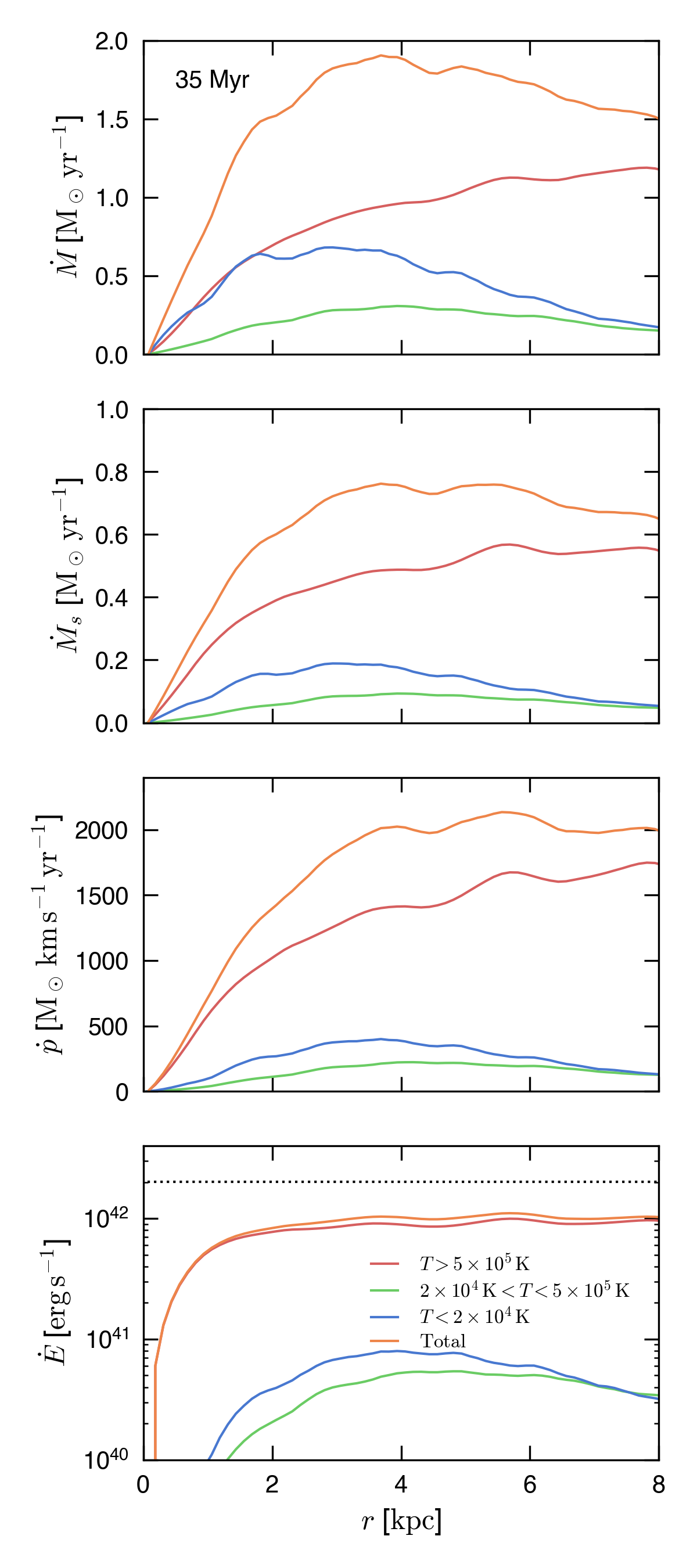}
    \caption{Radial outflow rates for mass, scalar mass, momentum, and total energy at 35 Myr. The outflow rates are measured in a bicone with a half opening angle of $30\degree$, and split by phase: hot ($T > 5\times10^5\,\mathrm{K}$, red line), intermediate ($2\times10^4\,\mathrm{K} < T < 5\times10^5\,\mathrm{K}$, green line), and cool ($T < 2\times10^4\,\mathrm{K}$, blue line). The orange line shows the sum. The dashed horizontal line in the energy panel shows the total energy injection rate normalized by the total scalar outflow rate that is measured in the cone.}
    \label{fig:fluxes_35}
\end{figure}

\begin{figure}
    \centering
    \includegraphics[width=\linewidth]{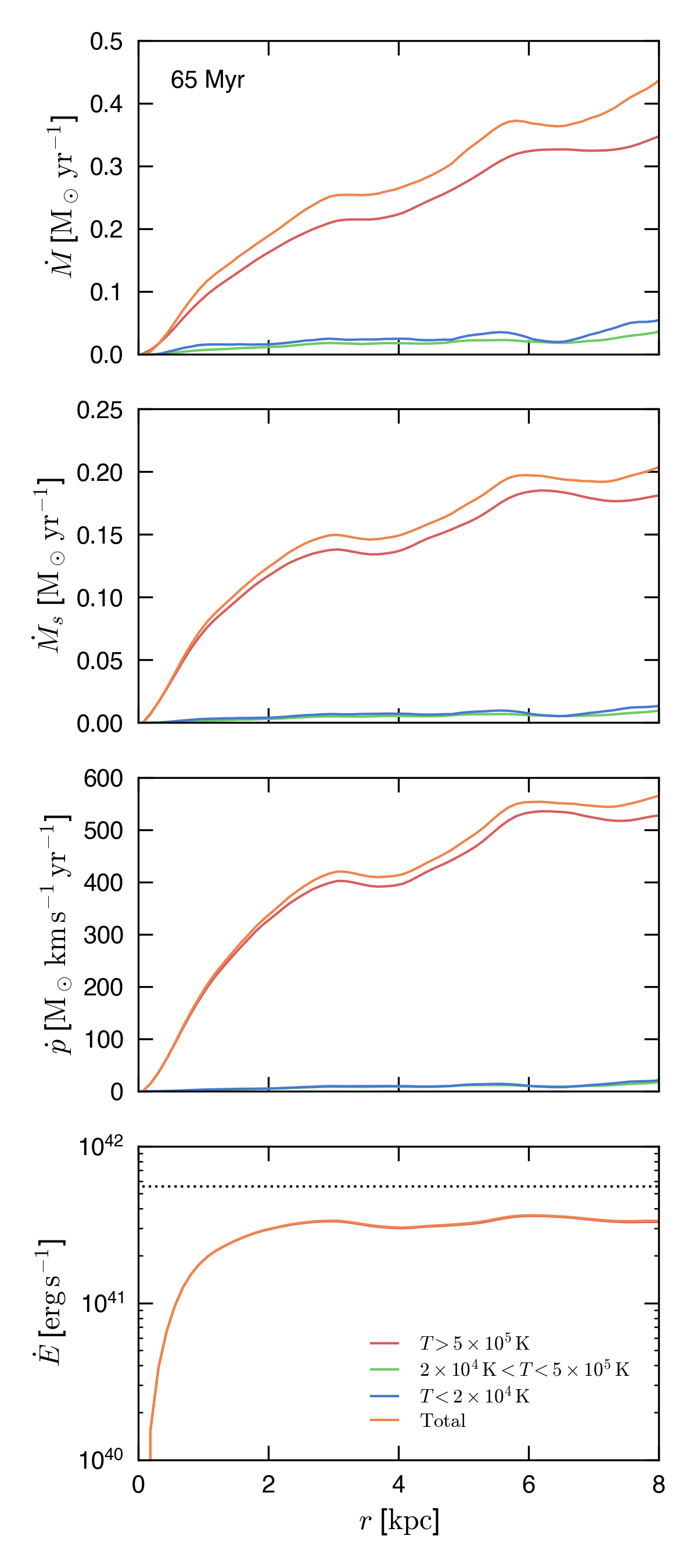}
    \caption{Radial outflow rates for mass, scalar mass, momentum, and energy at 65 Myr. All lines are as in Figure~\ref{fig:fluxes_35}.}
    \label{fig:fluxes_65}
\end{figure}

We show the radial outflow rates in the biconical region for mass, scalar mass, momentum, and energy at 35~Myr in Figure~\ref{fig:fluxes_35}, and at 65~Myr in Figure~\ref{fig:fluxes_65}. For the momentum outflow rate, we only include the kinetic component, that is $\dot{p} = \sum (M v^2_{r}) / \Delta r$. The energy outflow rate is calculated using the total energy, and including both the kinetic and the thermal component, $\dot{E} = \sum (E v_{r}) / \Delta r$, where $E = \rho(\frac{1}{2}v^2 + e_\mathrm{th})$, and $e_\mathrm{th}$ is the specific internal energy\footnote{For both the momentum and energy outflow rates, we are not including the additional pressure contribution to the flux. Hence, the term ``flux" as applied in this section is for convenience, but all rates should be assumed to be calculated as defined here.}. The outflow rates are split according to the temperature of the gas, into hot ($T > 5\times10^5\,\mathrm{K}$), intermediate ($2\times10^4\,\mathrm{K} < T < 5\times10^5\,\mathrm{K}$), and cool ($T < 2\times10^4\,\mathrm{K}$) phases. Including the intermediate phase allows us to better understand the amount of mixing that is happening in the outflow, since much of the gas in this temperature range has a relatively short cooling time and is therefore a transient phase (see Section~\ref{sec:histograms}). We scale down the vertical axes by a factor of 4 in Figure~\ref{fig:fluxes_65} relative to Figure~\ref{fig:fluxes_35} to better compare given the lower ``star formation rate'' at late times - $\dot{M}_\mathrm{SFR, early} = 20\,\mathrm{M}_\odot\,\mathrm{yr}^{-1}$ versus $\dot{M}_\mathrm{SFR, late} = 5\,\mathrm{M}_\odot\,\mathrm{yr}^{-1}$. Accounting for that normalization, we see that the total outflow rates are fairly consistent between the early and late stages of the outflow, but the intermediate and cool fractions are much smaller at lower outflow rates.

Figures~\ref{fig:fluxes_35} and \ref{fig:fluxes_65} also demonstrate some natural relationships for the outflow. Most of the energy and momentum are carried by the hot phase of the wind at all times, as would be expected for an energy-driven wind model. At small radii and early times, when the average density in the wind is high, there are comparable amounts of mass in both the hot and the cool phases (particularly if we include the intermediate temperature gas with the cool). There are times when the mass outflow rate in the cool gas is larger than that of the hot gas (not shown in these figures), but they are always within a factor of $\sim2$ in our simulations. That is to say, the mass outflow rate is never \textit{dominated} by the cool gas. In each of the energy outflow rate panels, we have added a dotted line showing the total energy available given our input energy injection rate and assuming no losses (so, $\alpha_\mathrm{eff} = 1$). At both early and late times, we see that the total energy outflow rate is below this line by a factor of $\sim2$. It is interesting to note that the total momentum outflow rate in the cone, despite representing only a fraction of the total outflow, is nevertheless within a factor of two of $L/c$ that would be expected for a freely expanding hot wind, which would be $\dot{p}\approx 4000\Msun\,\mathrm{km}\,\mathrm{s}^{-1}\,\mathrm{yr}^{-1}$.

We can also use these outflow rates to assess the validity of the time steady assumption that went into our analytic model. At 35 Myr, the total outflow rates are all fairly flat as a function of radius, outside of $R = 2 - 3\,\mathrm{kpc}$. Thus, we interpret the outflow at this time as an approximately time-steady flow. The flatness does depend on phase, however. While the mass outflow rate of the hot phase is increasing with radius, the cool phase is decreasing, and the intermediate temperature gas remains roughly constant. This is a clear indication that mass is being transferred from the cool phase to the hot as a result of mixing between the phases. In addition, the slight decrease in the total mass outflow rate between $R = 4 - 8$ may indicate the presence of a fountain flow in the cool phase, as lower velocity gas drops out at higher radii \citep[see e.g.][]{Kim18}.

Returning to the momentum outflow rates in Figure~\ref{fig:fluxes_35}, we see that although the total momentum outflow rate is roughly constant outside of $R\sim4\,\mathrm{kpc}$, the momentum outflow rate in the cool phase is actually decreasing. This may seem counter-intuitive, since the profiles in Figure~\ref{fig:profiles_cool} showed that the velocity of the cool gas is increasing as a function of radius, and we have attributed this increase to a transfer of momentum from the hot phase. In addition, the energy outflow rate in the hot phase stays roughly constant outside of $R\sim2\,\mathrm{kpc}$, which would seem to call into question an explanation in which the mixed intermediate temperature gas radiates away its thermal energy and adds mass to the cool phase with an increased velocity \citep[i.e.][]{Gronke18}. In that case, we might expect the hot phase to show a decreasing energy flux with radius, reflecting this lost energy.

\begin{figure}
    \centering
    \includegraphics[width=\linewidth]{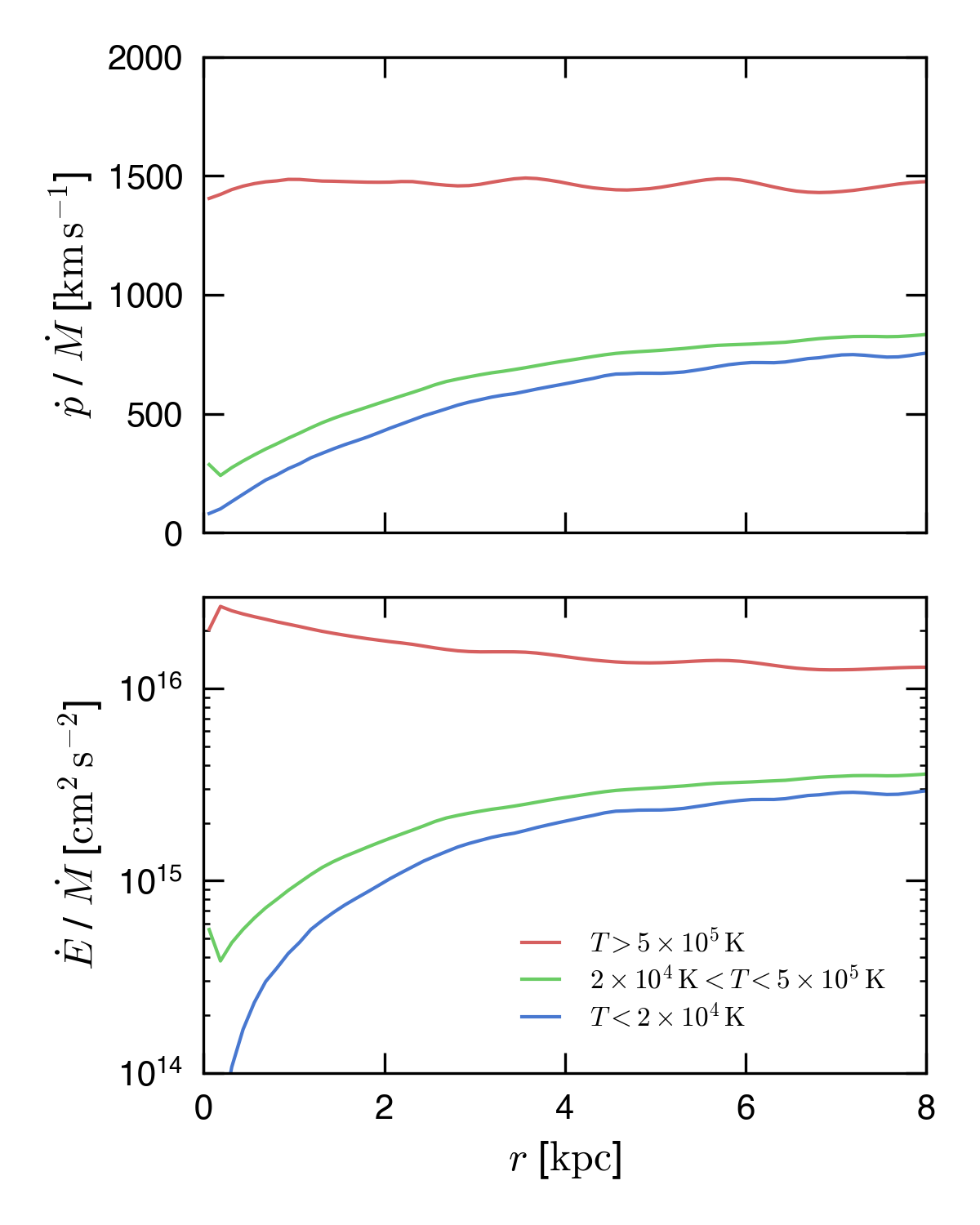}
    \caption{Radial outflow rates for momentum and energy at 35 Myr, normalized by mass outflow rates in each phase.}
    \label{fig:fluxes_norm}
\end{figure}

To shed some light on the matter, we show in Figure~\ref{fig:fluxes_norm} versions of the momentum and energy outflow rates normalized by the mass outflow rate in each phase. This corresponds to a velocity as a function of radius in the top panel, and a specific total energy ($e_\mathrm{T} = \frac{1}{2}v^2 + e_\mathrm{th}$) as a function of radius in the bottom panel. With the total mass normalized out, we see that the gas in each phase behaves exactly as we would expect given the mixing model we have outlined. While the velocity in the hot phase stays roughly constant as a function of $r$, the velocity in the intermediate and cool phases is consistently rising. The specific energy of the cool and intermediate phases is also rising, reflecting the higher kinetic energy at larger radii. Meanwhile, the specific energy in the hot phase decreases slightly as a function of radius, reflecting the that low energy cool gas is being mixed in with the hot, lowering the average specific energy of the hot gas as a function of radius. The energy loss from cooling by the hot phase is fractionally quite small, such that the outward decrease in the specific energy is mainly due to the outward mass increase of the hot, and thus we do not see a decrease in Figure~\ref{fig:fluxes_35}.

%\subsection{Comparison to the Analytic Model}\label{sec:comparison}

%In Section~\ref{sec:analytics}, we described an analytic model that relates the radial profiles of Section~\ref{sec:profiles} to the radial fluxes of Section~\ref{sec:fluxes}. Here, we test how well this model actually reproduces those fluxes. In particular, we compare the mass and energy fluxes calculated from the profiles, to those measured in the simulation. Before making the comparison, we smooth both the profiles and the fluxes using a third order Savitzky-Golay filter to decrease the noise when calculating e.g. derivatives. Figure~\ref{fig:analytic_comparison} shows the results. Both the mass and energy fluxes are reproduced well by the profiles, demonstrating that the analytic model is capturing the relevant physics.

%\begin{figure}
%    \centering
%    \includegraphics[width=\linewidth]{mass_flux_comparison.png}
%    \includegraphics[width=\linewidth]{Energy_flux_comparison.png}
%    \caption{Comparison between the mass and energy fluxes measured in the simulations, and those given by the analytic model. For the mass flux, we use the density-weighted profiles shown in Figure~\ref{fig:profiles_hot}. For the energy flux, we use volume-weighted profiles.}
%    \label{fig:analytic_comparison}
%\end{figure}

\subsection{Mass and Energy Loading}\label{sec:loading}

\begin{figure}
    \centering
    \includegraphics[width=\linewidth]{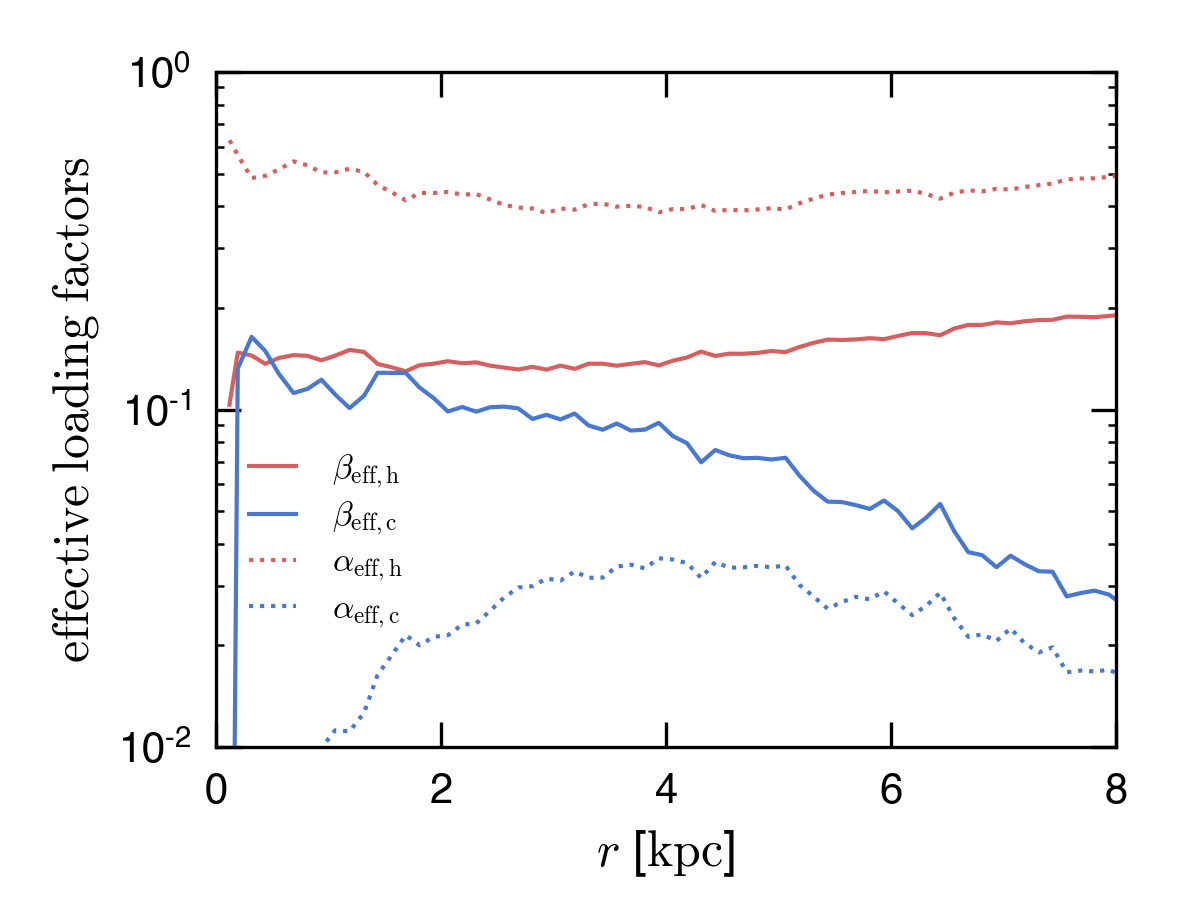}
    \caption{Effective mass and energy loading factors as a function of radius for the hot and the cool phases of the wind at 35 Myr. Loading factors are measured within the same $30\degree$ bicone used to measure the profiles.}
    \label{fig:loading_35}
\end{figure}

\begin{figure}
    \centering
    \includegraphics[width=\linewidth]{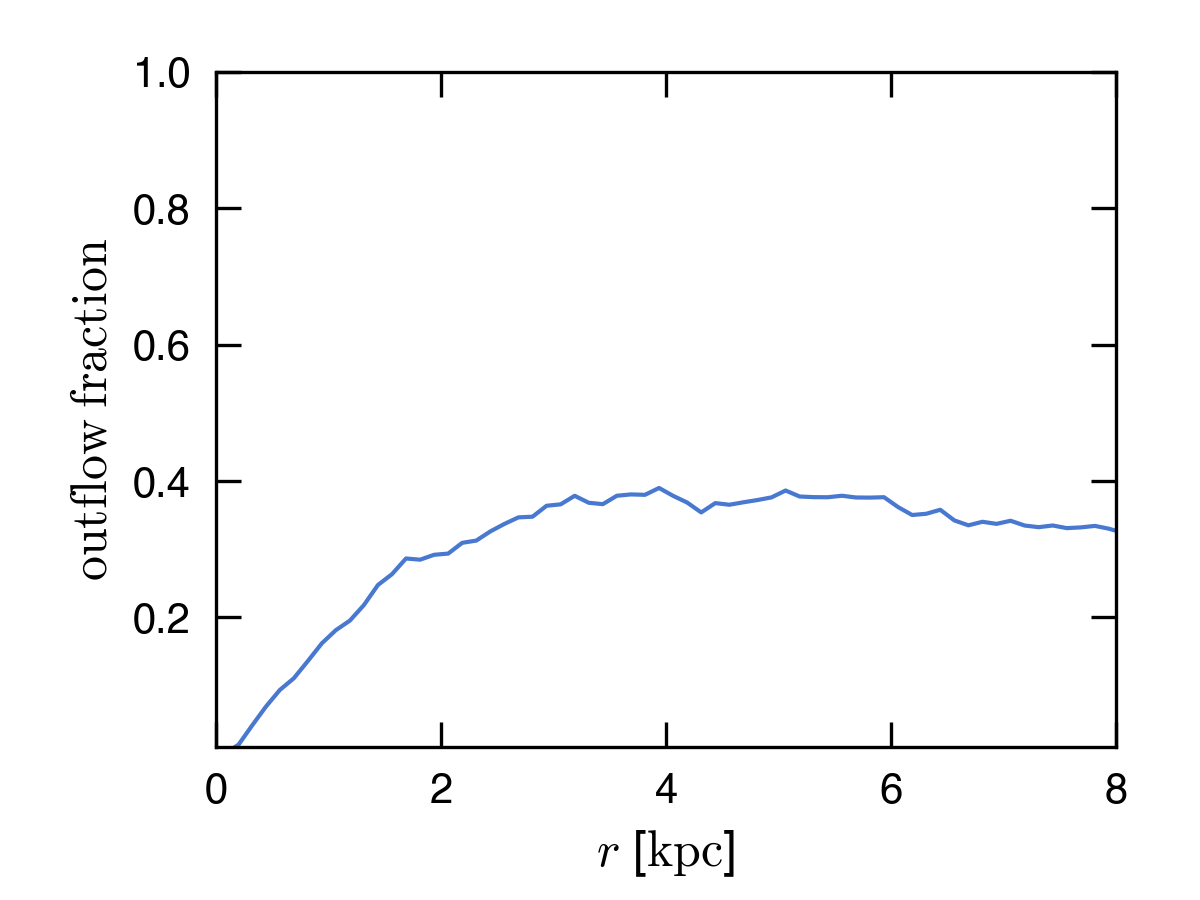}
    \caption{Fraction of the total injected mass that is captured in our conical selection region.}
    \label{fig:outflow_fraction}
\end{figure}

Using the profiles measured in Section~\ref{sec:profiles}, and the relationships defined in Section~\ref{sec:analytics}, we can calculate the effective mass and energy loading in the different phases of the outflow. Using Equation~\ref{eqn:eff_beta}, we can straightforwardly calculate the effective mass loading in the hot outflow as the injected mass loading factor ($\beta_\mathrm{inj}\sim 0.1$) divided by the passive scalar, $s$, and multiplied by the factor $\dot{M}_{s,\mathrm{net,h}} / \dot{M}_\mathrm{inj}$. Because we have measured the outflow rates in a cone, we must ``correct" the denominator, $\dot{M}_\mathrm{inj}$, which represents the total injected mass rate (all hot), to the total mass injected in the cone, which is just $\dot{M_{s}}$ (all phases). This correction factor is plotted in Figure~\ref{fig:outflow_fraction}. Carrying out this calculation, we find $\beta_\mathrm{eff,h} = 0.17$, approximately a factor of six below the star formation rate. We note, however, that $\beta_\mathrm{eff,h}$ ranges from 0.1 at small radii to 0.2 at 8 kpc - a result of the additional mass loading of the hot phase taking place as some of the cool gas is mixed in. We plot these values as a function of radius in Figure~\ref{fig:loading_35}. Similarly, we can calculate the effective energy loading in the hot outflow using equation~\ref{eqn:eff_alpha1} or equation~\ref{eqn:eff_alpha}, again corrected for the amount of mass injected into the cone. Using equation~\ref{eqn:eff_alpha1}, we find a roughly constant value of energy loading within the hot phase, $\alpha_\mathrm{eff} = 0.4-0.5$, implying that $\sim50-60\%$ of the input energy has been radiated away in the outflow\footnote{Technically, this is the fraction of the energy in the hot phase, but because most of the energy is contained in the hot phase, it is a decent approximation for the radiative losses as a whole.}. While this implies a robust energy-driven wind, we caution against over-interpreting the exact measured values, since they may be sensitive to the cluster input scheme, in the sense that having only a few massive clusters minimizes the interaction of supernovae with the ISM, and thus is likely to maximize the value of $\alpha_\mathrm{eff}$ while potentially minimizing $\beta_\mathrm{eff}$. We will explore this possibility in future work. 

\begin{figure}
    \centering
    \includegraphics[width=\linewidth]{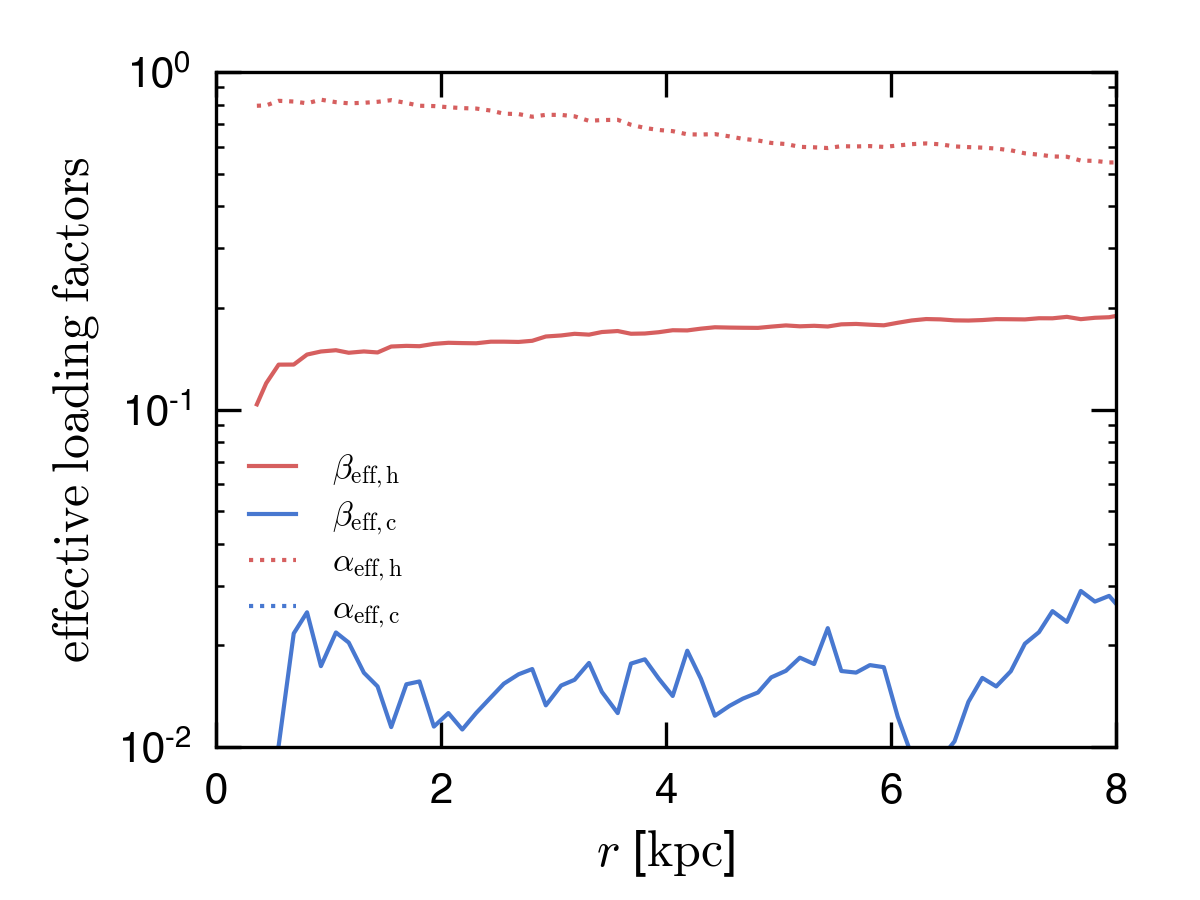}
    \caption{Effective mass and energy loading factors as a function of radius for the hot and the cool phases of the wind at 65 Myr. Loading factors are measured within the $30\degree$ cone used to measure the profiles.}
    \label{fig:loading_late}
\end{figure}

Following the same procedure (applying Equations~\ref{eqn:eff_beta} and \ref{eqn:eff_alpha} corrected for the mass injected into the cone), we also calculate the effective mass and energy loading for the cool phase, as shown in Figure~\ref{fig:loading_35}. The loading factors for the hot and cool phases demonstrate the same relationships seen in the profiles and outflow rates - at small radii, a significant fraction of the outflow is in the cool phase, but at larger radii, that gas has mostly been incorporated into the hot phase or dropped out of the flow. As a result, the mass loading factor for the hot phase increases, while the cool phase decreases. As demonstrated by $\beta_\mathrm{eff,c}$, at the maximum point around $r = 4\,\mathrm{kpc}$, the cool phase has coupled a few percent of the total energy available.

Given these mass and energy loading factors, we can now explain why this CGOLS simulation does not lead to the large-scale cooling of the hot wind seen in \citetalias{Schneider18b}. Following \cite{Thompson16}, we estimated the cooling radius of the hot wind as
\begin{equation}
r_\mathrm{cool} \simeq 4\,\mathrm{kpc} \frac{\alpha^{2.13}}{\beta^{2.92}}\mu^{2.13}R_{0.3}^{1.79}\left(\frac{\Omega_{4\pi}}{\dot{M}_{\mathrm{SFR},10}}\right)^{0.789}.
\label{eqn:r_cool}
\end{equation}
With $\alpha = 0.5$ and $\beta = 0.2$, and using $R = 1\,\mathrm{kpc}$ for the injection radius, we calculate $r_\mathrm{cool} = 170\,\mathrm{kpc}\,\Omega_{4\pi}^{0.789}$. While we have demonstrated that $\Omega$ may be significantly smaller than $4\pi$, a conservative lower limit of $\nicefrac{1}{8}$ still gives $r_\mathrm{cool} = 33\,\mathrm{kpc}$, well outside the bounds of our simulation box. \cite{Thompson16} also estimate a minimum $\beta$ below which cooling of the hot wind would not take place at any radius (see their Eqn. 7), which with these same values would be $\beta = 0.17$. Given that $\Omega$ is likely underestimated in this case, the volume-filling hot wind in this simulation is therefore unlikely to cool. Nevertheless, we note that this cooling radius depends strongly on the exact values of $\alpha$, $\beta$, and $\Omega$. Using the same $\beta$ and $\Omega$, but a slightly lower $\alpha = 0.1$, $r_\mathrm{cool} \sim 1\,\mathrm{kpc}$, so we do not consider this work to \textit{rule out} cooling of the hot wind.

%Given that we know what the input mass and energy input rates were for the simulation, we can compare these profile-based estimates for the effective mass and energy loading with those we actually measured in Section~\ref{sec:fluxes}. At a radius of $r = 5\,\mathrm{kpc}$, the total mass flux in the hot phase is $1\,\Msun\,\mathrm{yr}^{-1}$, and the energy flux at the same radius is $9\times10^{41}\,\mathrm{erg}\,\mathrm{s}^{-1}$. Correcting for the rest of the sphere, we would recover a mass flux of $7.5\,\Msun\,\mathrm{yr}^{-1}$, and an energy flux of $7\times10^{42}\,\mathrm{erg}\,\mathrm{s}^{-1}$. At 35 Myr, our average input mass injection rate is $\dot{M} = 2\,\Msun\,\mathrm{yr}^{-1}$, and the energy injection rate is $\dot{E} = 6\times10^{42}\,\mathrm{erg}\,\mathrm{s}^{-1}$. Thus, it would seem that our estimated effected values are too high, because $\alpha$ cannot be greater than 1. However, as noted in the previous section, the outflow is not in fact spherical. If we measure the total fluxes within the full sphere, we recover a total mass flux of $5\,\Msun\,\mathrm{yr}^{-1}$, and an energy flux of $2.5\times10^{42}\,\mathrm{erg}\,\mathrm{s}^{-1}$, which match the effective mass and energy loading calculated using the profiles quite well.

In addition to allowing us to calculate effective loading factors within the cone, the correction factor $\dot{M}_s / \dot{M}_\mathrm{inj}$ also gives us an estimate for what fraction of the total outflow is contained within the cone. This factor steadily increases at small radii, plateauing at around 0.35 at $r\sim 3-4\,\mathrm{kpc}$. This means that roughly $\nicefrac{1}{3}$ of the total injected cluster mass is escaping within the biconical region, more than twice the rate that would be expected in a spherical outflow, implying that the disk is having a significant collimating effect. Of course, this analysis would not capture the impact of a higher rate of mass loading outside the biconcial region, or a lower energy loading. In fact, when we look at the mass, momentum, and energy outflow rates outside of the bicone, we do find higher values of $\beta_\mathrm{eff}$, particularly for the cool phase at small radii. However, we also find that the total mass outflow rate falls from $\dot{M} = 6\,\Msun\,\mathrm{yr}^{-1}$ to $\dot{M} = 5\,\Msun\,\mathrm{yr}^{-1}$ between $r = 1 - 5\,\mathrm{kpc}$ within the full sphere, implying that the effects of a cool fountain are perhaps more significant at larger angles, making the large angle contributions to the total mass outflow rate less important at larger radii.

\subsection{Phase Diagrams}\label{sec:histograms}

In addition to looking at properties of the outflow as a function of radius, we can use 2D histograms to better characterize the relationships between different physical properties. In this section, we explore how density, velocity, and temperature relate to one another.

\begin{figure}
    \centering
    \includegraphics[width=\linewidth]{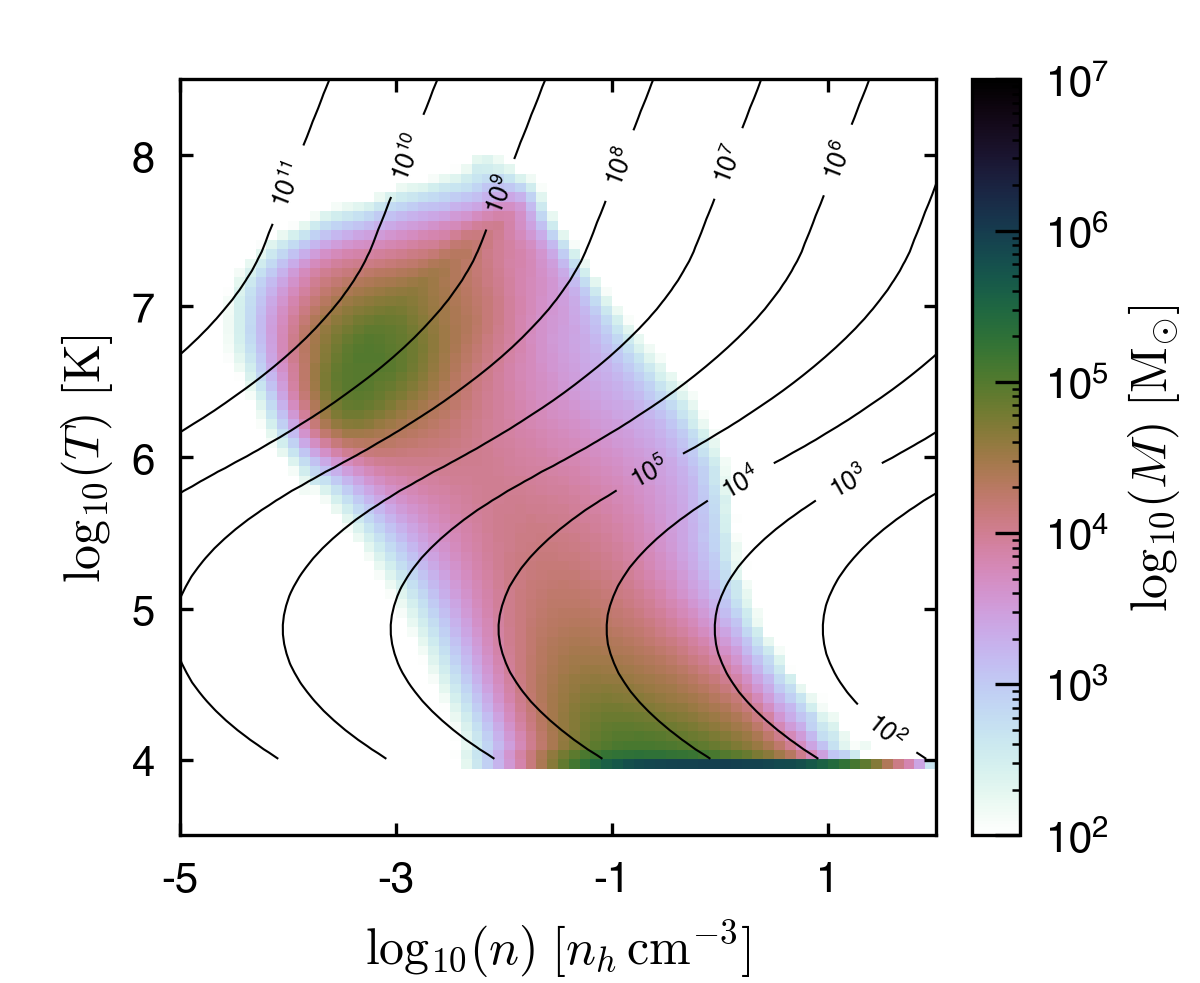}
    \caption{Mass-weighted phase diagram for gas within the biconical outflow region at 35 Myr. Contours label the approximate cooling time of the gas, in years.}
    \label{fig:nT_35}
\end{figure}

In Figure~\ref{fig:nT_35} we show a phase diagram of the gas within the biconical region during the high star formation state. The bins are weighted by total mass, to highlight the regions where the bulk of the mass resides. We additionally exclude the regions within 0.5 kpc of the midplane, to avoid including pure disk gas or the interior of a cluster in our analysis. The histogram shows two clear peaks, demonstrating the primarily two-phase nature of the outflow. Integrating these regions, we find that there is slightly more total mass in the cool phase than the hot - $2.5\times10^7\,\Msun$ and $1.7\times10^7\,\Msun$, respectively. The ratios reverses during the low star formation state, however. The same integration at 65 Myr yields $3.1\times10^6\,\Msun$ and $4.6\times10^6\,\Msun$ for cool vs hot. These total values are consistent with there being about 4 times more outflowing gas during the high star formation state (SFR = $20\,\Msun\,\mathrm{yr}^{-1}$) than the low (SFR = $5\,\Msun\,\mathrm{yr}^{-1}$), indicating a roughly constant mass-loading factor, as indeed is seen in Figure~\ref{fig:loading_late}.

\begin{figure}
    \centering
    \includegraphics[width=\linewidth]{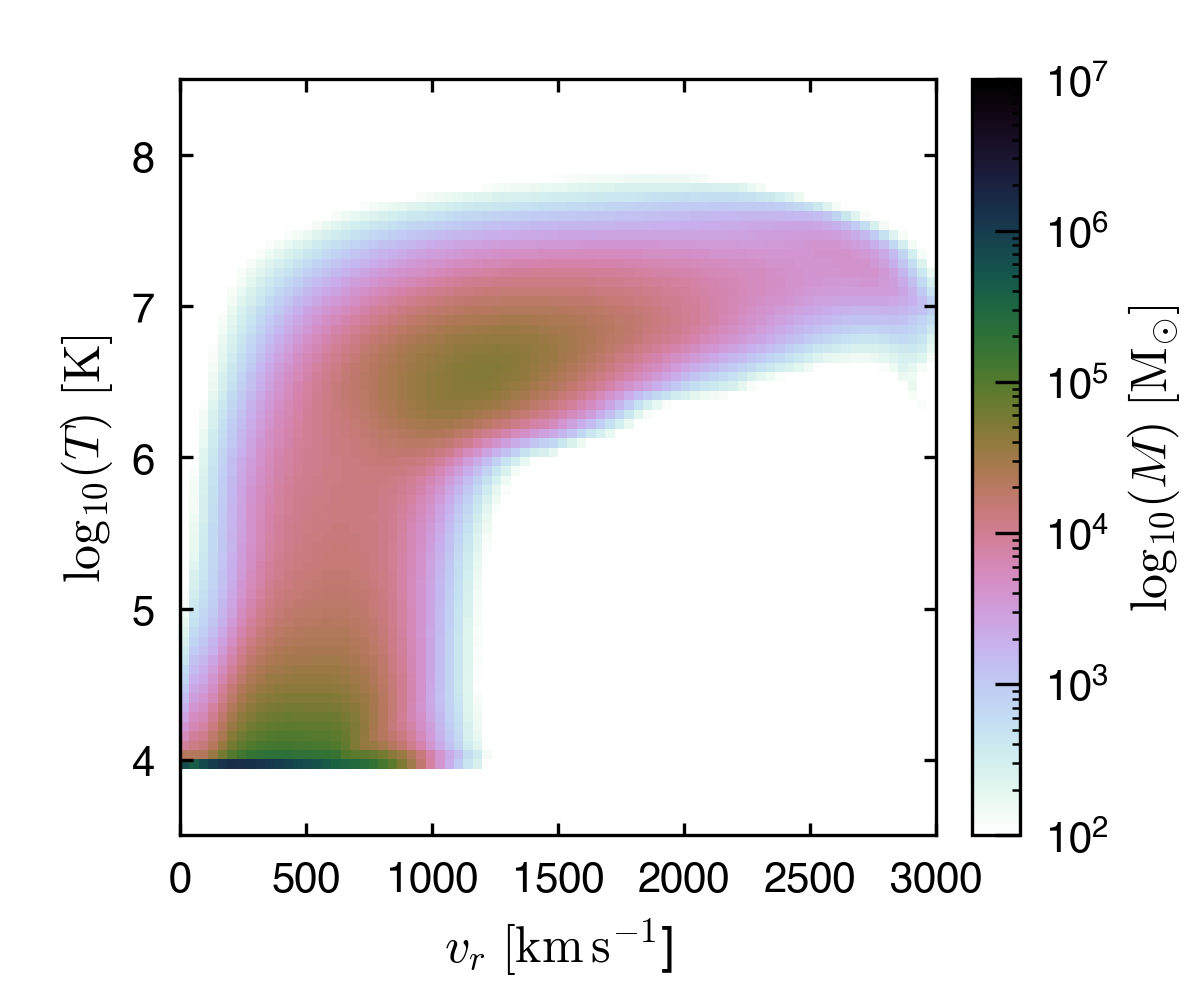}
    \caption{Mass-weighted velocity temperature histogram for gas within the biconical outflow region at 35 Myr.}
    \label{fig:vT_35}
\end{figure}

We can estimate the cooling time of the gas as
\begin{equation}
    t_\mathrm{cool} = \frac{k_\mathrm{B}T}{n \Lambda(n, T)},
\end{equation}
where $\Lambda(n, T)$ is the density and temperature dependent cooling function, measured in $\mathrm{erg}\,\mathrm{s}^{-1}\,\mathrm{cm}^{3}$. We show these cooling times plotted over the phase diagram in Figure~\ref{fig:nT_35}. While this demonstrates that the hot gas in the wind has very long cooling times, and the temperature floor of $T = 10^4\,\mathrm{K}$ means that cool gas in our simulations has effectively infinite cooling time, there exists significant mass in gas at intermediate temperatures ($2\times10^4\,\mathrm{K} < T < 5\times10^5\,\mathrm{K}$) that is in an interesting regime. Given an approximate number density of $n = 10^{-1}\,\mathrm{cm}^{-3}$, and temperature at the peak of the cooling curve, $T \approx 10^5\,\mathrm{K}$, the approximate cooling time (assuming solar metallicity) is $t_\mathrm{cool} \sim 10\,\mathrm{kyr}$. For gas traveling at the hot gas speed of $v \geq 1000\,\mathrm{km}\,\mathrm{s}^{-1}$, this is comparable to the dynamical time of the wind, 
\begin{equation}
    t_\mathrm{adv} = 10\,\mathrm{kyr} \left[\frac{r}{10\,\mathrm{kpc}}\right]  \left[\frac{v}{1000\,\mathrm{km}\,\mathrm{s}^{-1}}\right]^{-1}.
\end{equation}
We therefore conclude that a significant fraction of the intermediate temperature gas (which tends to be moving at $v < 1000\,\mathrm{km}\,\mathrm{s}^{-1}$, see Figure~\ref{fig:vT_35}) can cool to $T \sim 10^4\,\mathrm{K}$ while in the outflow. In particular, the gas that is most prone to cooling is the highest density, lowest temperature portion of the intermediate temperature gas that is created by mixing and shocks. The flat nature of the $\dot{M}_\mathrm{int}$ flux seen in Figures~\ref{fig:fluxes_35} and \ref{fig:fluxes_65} is an indication that this mixing is taking place at all radii, and the mixing rate is roughly balanced by the cooling rate. We expect this to hold as long as there is $T \sim 10^4\,\mathrm{K}$ gas present at that radius.

In addition to the relationship between density and temperature, we can also investigate how velocity and temperature correlate in the outflowing gas. In Figure~\ref{fig:vT_35}, we show a mass-weighted velocity temperature histogram. Again, the two peaks highlight that the bulk of the gas is either in the hot, fast-moving wind ($v > 1000\,\mathrm{km}\,\mathrm{s}^{-1}$, $T > 10^6\,\mathrm{K}$), or a cooler, slower phase ($v < 800\,\mathrm{km}\,\mathrm{s}^{-1}$). By mass, the bulk of the cool gas has velocities less than $500\,\mathrm{km}\,\mathrm{s}^{-1}$, but there is a substantial tail with velocities all the way up to $1000\,\mathrm{km}\,\mathrm{s}^{-1}$, as was also demonstrated in the velocity profiles shown in Figure~\ref{fig:profiles_cool}. This stands in contrast to our previous work using idealized cloud-wind simulations \citep{Schneider17}, where only high temperature gas attained high velocities. However, we note that these simulations are lower resolution than that work, and hence there is more numerical diffusion. We address this discrepancy in Section~\ref{sec:discussion}.

\begin{figure}
    \centering
    \includegraphics[width=\linewidth]{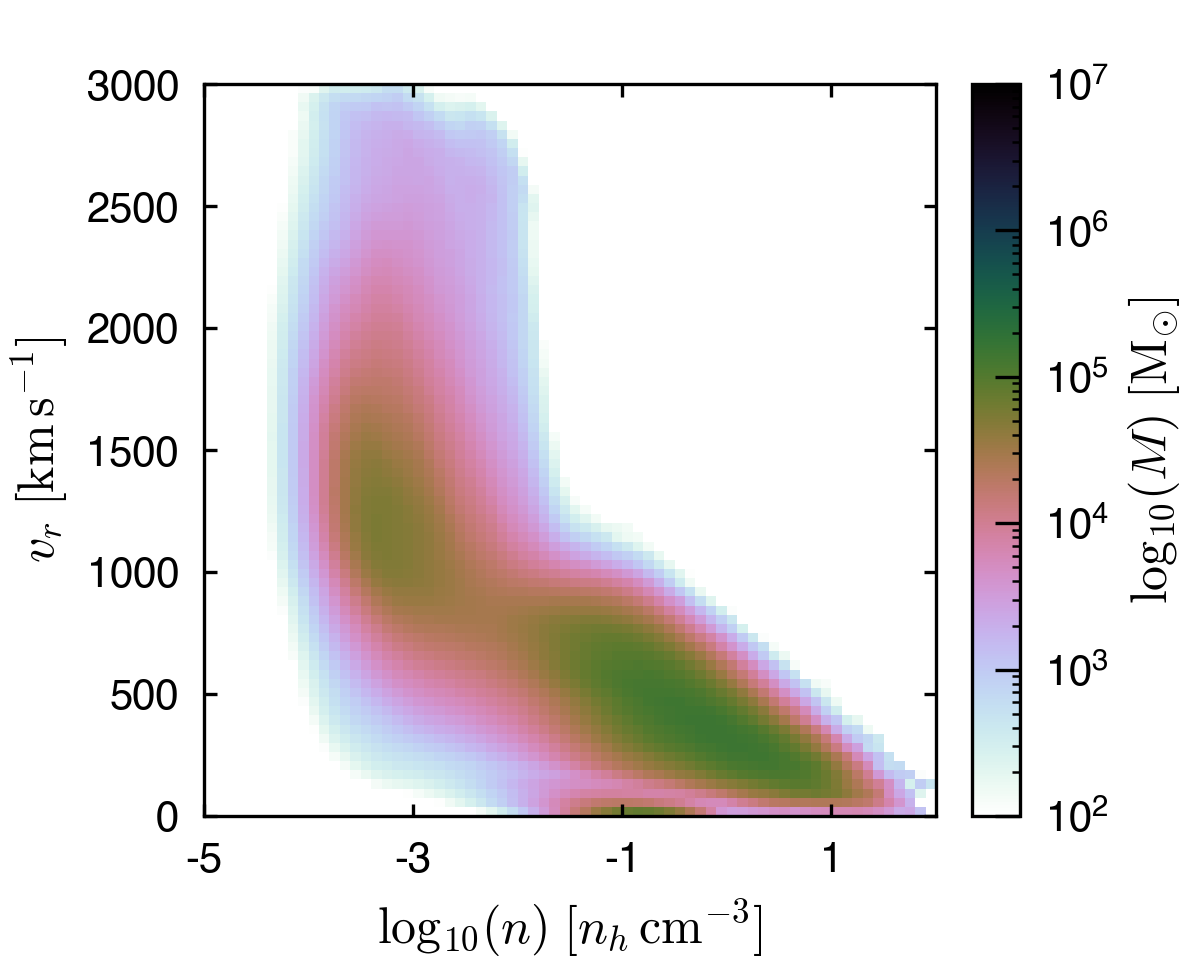}
    \caption{Mass-weighted density velocity histogram for gas within the biconical outflow region at 35 Myr.}
    \label{fig:nv_35}
\end{figure}

In Figure~\ref{fig:nv_35}, we show a mass-weighted density velocity histogram, again at 35 Myr. In addition to the correlation between velocity and radius demonstrated in Figure~\ref{fig:profiles_cool}, this figure shows that there is also a correlation between density and velocity. In other words, lower density gas appears to be more effectively accelerated, which is certainly consistent with intuition. The sharp cutoff for high velocity ($v > 1000\,\mathrm{km}\,\mathrm{s}^{-1}$) gas at $n\sim10^{-2}\,\mathrm{cm}^{-3}$ mirrors the cutoff in density for cool gas in Figure~\ref{fig:nT_35}, possibly indicating that this is the maximum velocity that can be achieved by cool gas in a starburst-driven outflow. We investigate the physical mechanism by which this cool gas is accelerated in the following subsection.

\subsection{Cool Gas Acceleration}\label{sec:acceleration}
A goal of this work was to assess the potential contributions of both mixing and ram pressure to acceleration of the cool gas, as those are the two primary theories for cool gas acceleration via hydrodynamic processes. We have excluded other potential sources of acceleration in this work (e.g. magnetic drag, radiation pressure, and cosmic ray pressure), in order to focus on the two processes most commonly invoked in the M82-like outflows we are studying here.

First, we consider the velocities we would expect for the cool gas, if \textit{all} of its momentum is gained via mixing with the hot phase of the wind. In this case, we would expect the momentum density of the cool gas, $\rho_\mathrm{c} v_\mathrm{c}$ to reflect the fraction of hot gas that has been mixed into the cool medium (and then cooled).  Since the cool gas initially has zero ``scalar mass" we can use  the scalar variable to determine the fraction of material in any cool cell that was previously hot.  From  Equation~\ref{eq:warmvelpred},
%\begin{equation}
%    \rho_\mathrm{w} v_\mathrm{w} = \rho_\mathrm{c, w} \frac{v_\mathrm{h}}{c_\mathrm{h}}.
%\end{equation}
%Using the fact that $\rho_\mathrm{c} = c \rho$, 
under the assumption of mixing-driven acceleration we obtain a linear relationship between the cool gas velocity and its scalar value:
\begin{equation}
    v_\mathrm{c} = \frac{v_\mathrm{h}}{s_\mathrm{h}} s_\mathrm{c}.
\label{eqn:cv}
\end{equation}
This relation is normalized by the velocity of the hot wind and its scalar value (this $s_\mathrm{h}$ does not have to be 1, because it reflects mixing of cool material into the hot gas at small radii).
%This expression can also be seen on the right-hand side of Equation~\ref{eqn:warm_momentum}. 

\begin{figure}
    \centering
    \includegraphics[width=\linewidth]{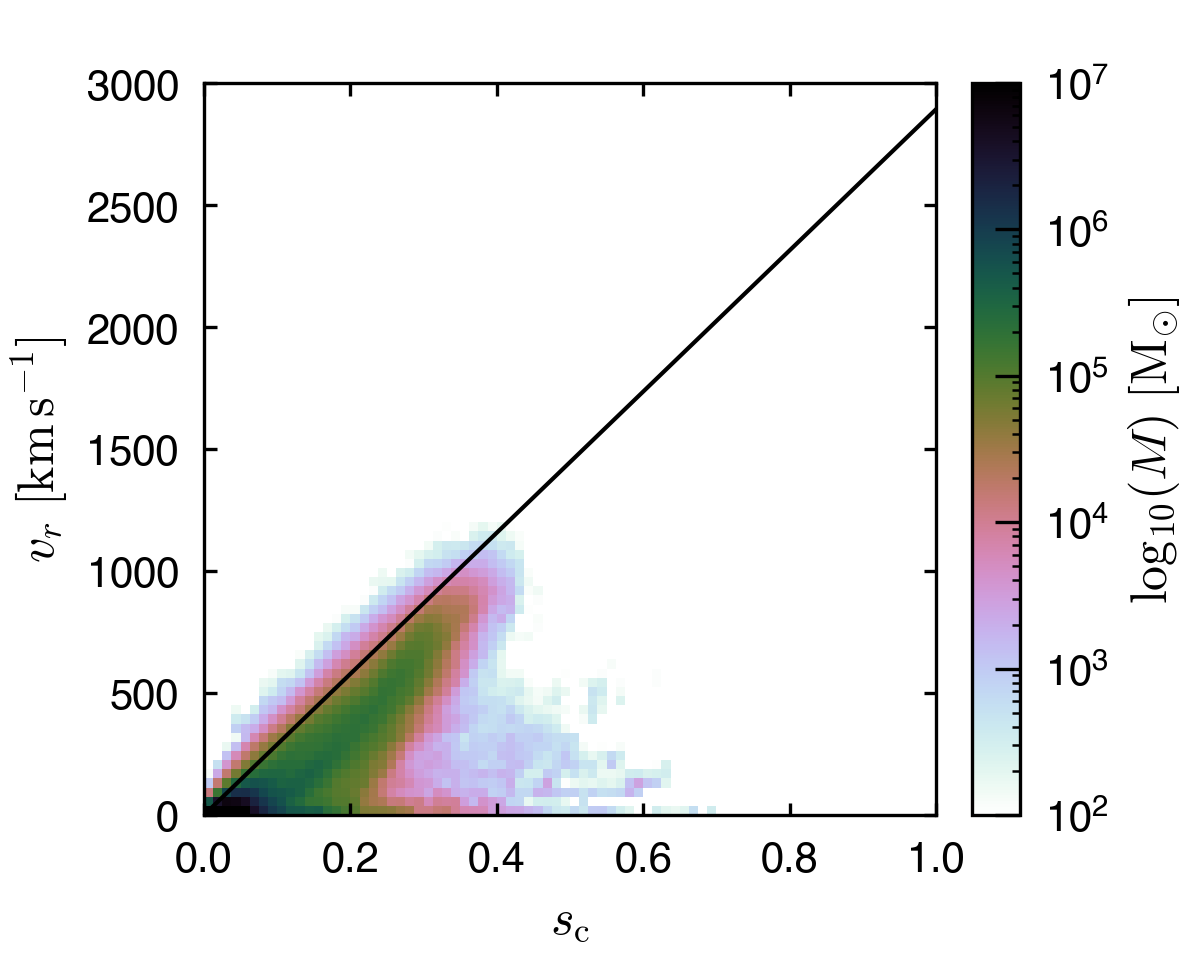}
    \caption{Mass-weighted histogram showing scalar value of the cool ($T < 2\times10^4$ K) gas versus its radial velocity. The black line shows the linear relationship expected if all of the momentum in the cool gas was transferred by mixing and cooling gas from the hot phase.}
    \label{fig:cv_cool}
\end{figure}

To test the validity of this hypothesis, we plot in Figure~\ref{fig:cv_cool} a mass-weighted histogram showing the radial velocity of the cool gas versus its scalar value, for all gas with $T < 2\times10^4$ within the same cone used for the earlier profiles and fluxes \cite[see also][]{Melso19}. We also plot the linear relationship expected given Equation~\ref{eqn:cv}. Because $v_\mathrm{h}$ and $s_\mathrm{h}$ are radially-dependent dependent quantities, we use their average values outside of 1 kpc to normalize the slope of the line, which yields a slope $v_\mathrm{h} / s_\mathrm{h} \approx 2900\,\mathrm{km}\,\mathrm{s}^{-1}$.

We see from Figure~\ref{fig:cv_cool} that the slope of the line is a reasonably good fit to the locus of the cool gas, particularly at larger values of $s_\mathrm{c}$, although the whole line is displaced upwards slightly. The fact that very little gas sits above the line demonstrates that ram pressure cannot have a dominant effect on the acceleration of the cool gas. Ram pressure without mixing would transfer momentum from the hot gas to the cool by accelerating it without raising its scalar value, which would move gas straight up on the plot. On the other hand, several effects could push gas below the line, including gravity, non-radial acceleration, and a lower value of the normalization $v_\mathrm{h} / s_\mathrm{h}$. Gravity is important only for the larger column-density clouds near the disk that are moving slowly - these also have the lowest value of $s$, and this is the part of the locus that lies the furthest from the line. Non-radial acceleration is expected given that the clusters are not distributed in a spherically-symmetric way, though they are concentrated at the center of the disk. Within the distribution radius, $R < 1\,\mathrm{kpc}$, mixing can accelerate gas in all directions, leading to an initial shift below the linear relation in Figure~\ref{fig:cv_cool} as scalar mass gets mixed in without producing radial acceleration. Lastly, the slope $v_\mathrm{h} / s_\mathrm{h}$ is not necessarily constant; indeed we see from Figure~\ref{fig:profiles_hot} that it should increase with radius. Because the histogram contains gas at all radii, we cannot directly represent this in the line plotted in Figure~\ref{fig:cv_cool}. That said, we see from Figure~\ref{fig:profiles_cool} that $s_\mathrm{c}$ is also increasing slightly with radius. In this case, we would expect the slope of the line to be shallower at low values of $s_\mathrm{c}$, and to steepen at higher values. This is consistent with the nature of the curve traced by the locus of the cool gas in Figure~\ref{fig:cv_cool}, which does appear flattest at small $s$ and increases in slope at larger $s$.

\subsection{Convergence}\label{sec:convergence}

\begin{figure}
    \centering
    \includegraphics[width=\linewidth]{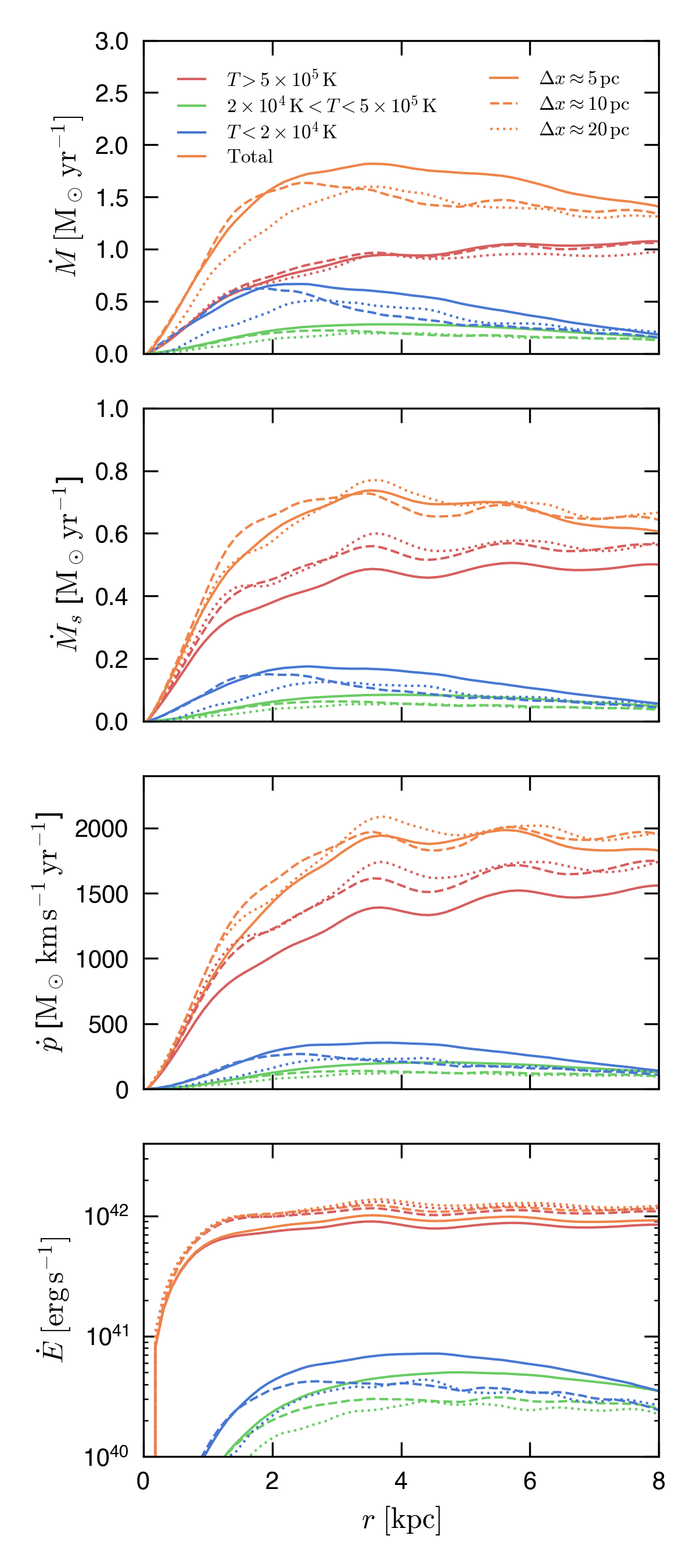}
    \caption{Mass, scalar mass, momentum, and energy outflow rates split by phase for simulations at three resolutions. The outflow rates are calculated in 0.25 kpc radial bins as in previous figures, but here we smooth over 3 binwidths and average from 33 - 37 Myr to mask out small scale time and spatial variability.}
    \label{fig:fl_convergence}
\end{figure}

Given that these results depend on the hydrodynamic mixing of gas with a large range of temperatures and densities, the question of whether they would hold at even higher resolutions is an important one. In order to assess convergence, we have run the same simulation at two additional resolutions, with $\Delta x \approx 10\,\mathrm{pc}$ and $\Delta x \approx 20\,\mathrm{pc}$. Because the input radius of the clusters themselves is $R = 30\,\mathrm{pc}$ in this simulation, we expect that they are marginally resolved even in the lowest resolution case. Any resolution dependence in the results is therefore likely to be a result of not capturing the mixing between the hot and cool gas in the disk and wind, as opposed to an unresolved feedback model.

We investigate the issue of convergence by comparing the mass, momentum, and energy outflow rates as a function of radius for each of the simulations. Figure~\ref{fig:fl_convergence} shows the resulting outflow rates for each simulation averaged between 33 - 37 Myr. In addition to averaging in time, we have additionally smoothed the radial fluxes using a third order Savitzky-Golay filter before time-averaging them to damp some of the small scale spatial variability and ease comparisons. 

\begin{figure}
    \centering
    \includegraphics[width=\linewidth]{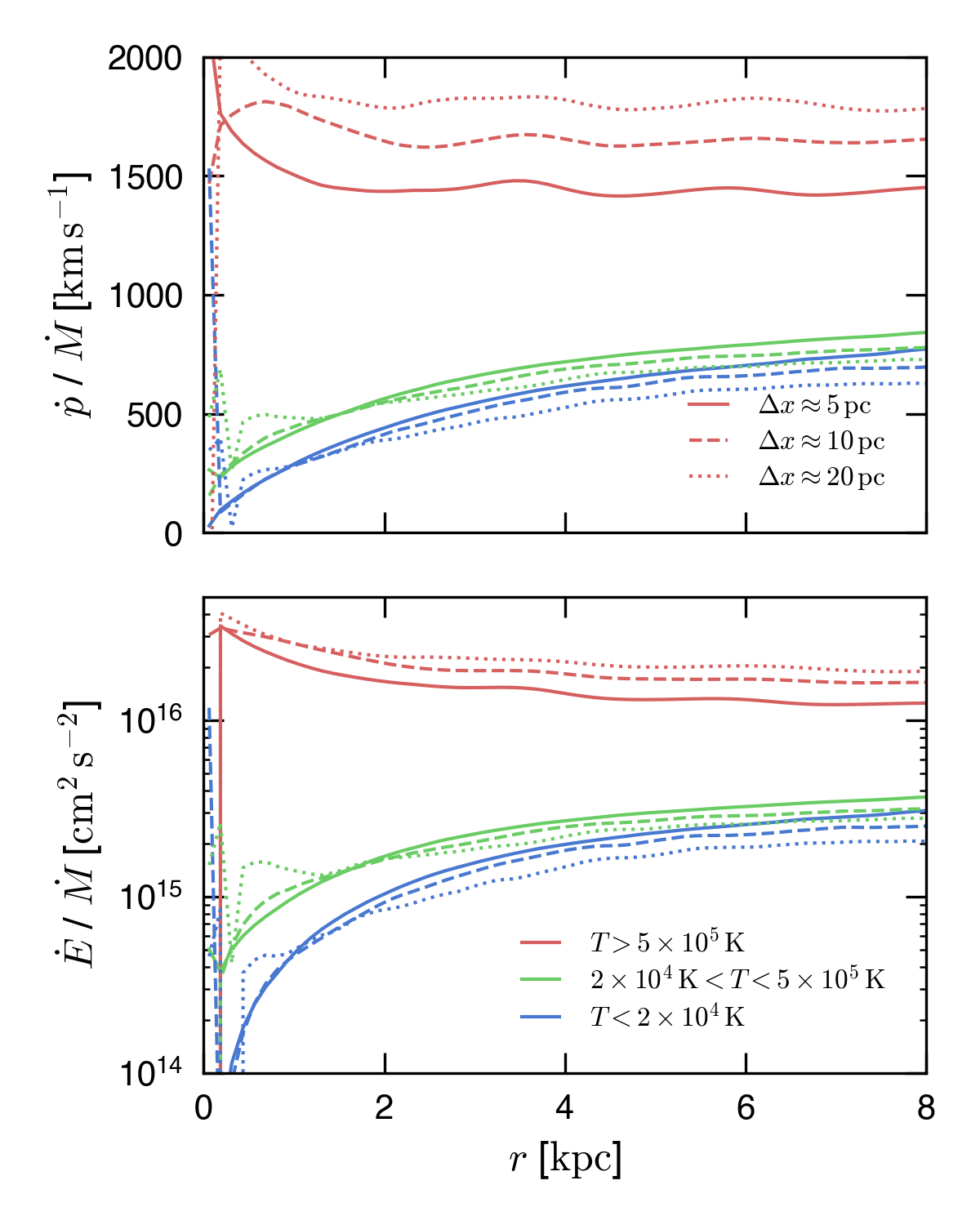}
    \caption{Radial outflow rates for momentum and energy normalized by the mass outflow rates. Rates are split by phase and shown for simulations at three resolutions. The outflow rates are calculated in 0.25 kpc radial bins as in previous figures, but here we smooth over 3 binwidths and average from 33 - 37 Myr to mask out small scale time and spatial variability.}
    \label{fig:f2_convergence}
\end{figure}

A primary result demonstrated in Figure~\ref{fig:fl_convergence} is that the mass, momentum, and energy outflow rates vary by less than 50\% for all phases at all radii between the three simulations. The biggest differences are in the cool gas, and the smallest are in the hot gas. The mass outflow rates do not show an obvious trend with resolution, though it is the case that the highest resolution simulation has the highest cool mass outflow rate. However, there are some potential trends that become apparent in the momentum and energy panels. At intermediate radii, the momentum and energy outflow rates in the cool and intermediate phases are higher in the $\Delta x = 5\,\mathrm{pc}$ simulation than at lower resolution. This indicates that momentum transfer from the hot to the cool gas is more efficient in the higher resolution simulation, as would be expected if that transfer is primarily a result of mixing. Such a relationship would \textit{not} be expected if the momentum transfer was primarily a result of ram pressure, since the lower resolution simulations tend to have larger clouds with more surface area perpendicular to the wind. At higher resolution, the cool clouds break up more, leading to less (perpendicular) surface area relative to total column density, and hence less efficient acceleration \citep[see][]{Schneider17}. The fact that the mass outflow rates for all phases are approximately equal at larger radii implies that although the details of the mixing process depend on the spatial resolution, the overall trend toward mass transfer from cool to hot gas as a function of radius holds.

We can see this relationship more clearly in Figure~\ref{fig:f2_convergence}, where we again normalize the momentum and energy outflow rates by the corresponding mass outflow rate. Here, the trend with resolution is obvious in both panels. As the resolution increases, the momentum in the hot phase decreases, while the momentum in the cooler phases increases. Similarly, the specific energy decreases faster in the hot phase at high resolution, and increases faster in the cool phase, indicating that the momentum transfer is more efficient at higher resolution, because mixing is more efficient.

%%%%%%%%%%%%%%%%%%%%%%%%%%%%%%%%%%%%%%%%%%%%%%%%%%%%%%%%%%%%%%%%%%%%%%%

\section{Discussion} \label{sec:discussion}

In this Section, we discuss some of the potential implications of these results and compare our findings to previous work. 

\subsection{Gas Velocity and Metallicity}

\begin{figure}[htb!]
    \centering
    \includegraphics[width=0.8\linewidth]{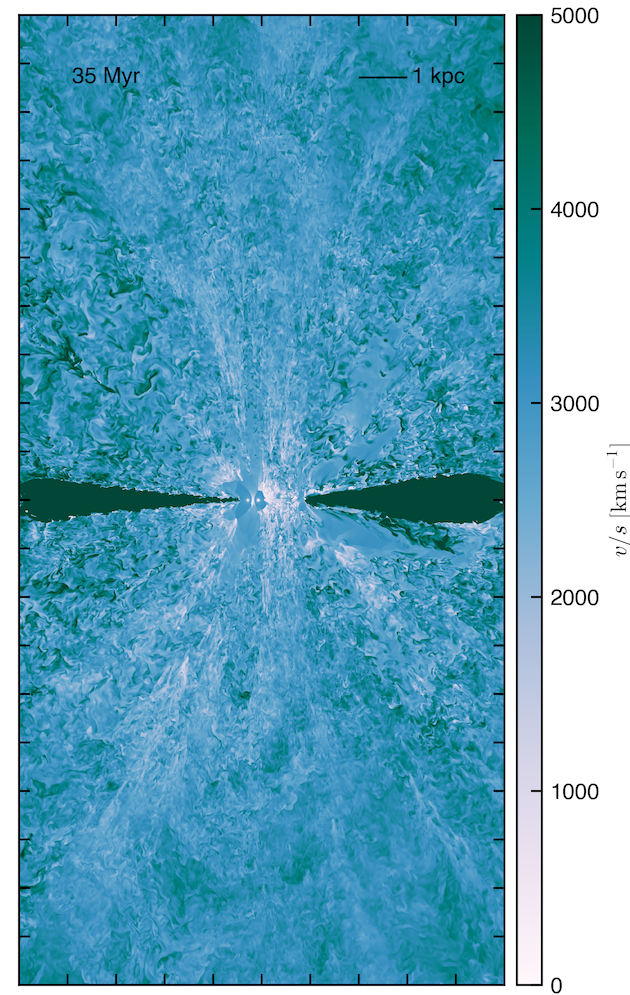}
    \caption{Velocity slice normalized by scalar mass at 35 Myr. The correspondence between scalar mass and velocity  leads to a relatively featureless outflow, and implies a correlation between metallicity and velocity in outflows.}
    \label{fig:metallicity}
\end{figure}

In Section \ref{sec:acceleration} we demonstrated using the scalar mass variable that acceleration of cool gas can be accounted for by mixing in some of the fast hot gas.  Those results also imply an expected relationship between the metallicity of the gas and its velocity. In Figure~\ref{fig:metallicity}, a velocity slice normalized by the associated scalar mass from the simulation at 35 Myr. While the correspondence between scalar mass and velocity is clear from Figure~\ref{fig:slices_35}, Figure~\ref{fig:metallicity} shows this directly - when normalized by the scalar, most of the large-scale velocity structure in the outflow is wiped out.

Because the injected gas in our model includes the directly deposited ejecta from supernovae, we expect the high scalar gas to be highly enriched in metals, while the disk gas has lower metallicity. Given a plausible metallicity for the disk and injected gas, it may thus be possible to back out a metallicity estimate for the cool gas observed in outflows, based on its velocity, as follows. In the case that all acceleration is due to mixing with hot gas that originally had velocity $v_\mathrm{h}$ and scalar value $s_\mathrm{h}$, it can be shown that
\begin{equation}
    Z(v) = \frac{Z_\mathrm{ej}s_{h}v + Z_\mathrm{ISM}v_h}{s_{h}v + v_h},
\label{eqn:Zv}
\end{equation}
where $Z_\mathrm{ej} = M_\mathrm{met,ej} / M_\mathrm{ej}$ is the metallicity of the supernova ejecta and $Z_\mathrm{ISM}$ is the original metallicity of the ISM. So, assuming that $Z_\mathrm{ej}$ and $Z_\mathrm{ISM}$ are known, we can use the roughly constant value of $s_h = 0.4$ from the simulation to estimate the metallicity of the hot gas:
\begin{equation}
    Z_h = \frac{Z_\mathrm{ej}s_h + Z_\mathrm{ISM}}{s_h + 1}.
\label{eqn:Zh}
\end{equation}
For the cool gas, one could use Equation~\ref{eqn:Zv} to obtain an estimate for its metallicity, based on its observed velocity $v$ and the observed or estimated values  of $v_h$ and $s_h$. 
%and assuming and $v_h = 1200\,\mathrm{km}\,\mathrm{s}^{-1}$ (also from the simulation). 
Alternatively, in the case that $Z_h$ is known from observations, then Equation~\ref{eqn:Zh} can be used to estimate $s_h$ in terms of $Z_h$, which can then be used in Equation~\ref{eqn:Zv} to estimate the metallicity of the cool gas.

%The linear relationship between velocity and scalar applies to hot gas in the simulation as well. An interesting trend has been reported in local galaxy halos, where super-solar absorbers tend to be located further out in the CGM \citep{Prochaska17}. This trend is naturally explained in the context of the outflow model presented here. Taking the scalar value as a proxy for gas metallicity, we see that high scalar channels are associated with high velocity channels (see Figure~\ref{fig:metallicity}). In other words, the less the hot gas has mixed, the faster it is going, and presumably, the farther it will propagate into the galaxy halo. This would lead naturally to the highest metallicity gas preferentially escaping to larger radii, leading to the observed inverse metallicity gradients.

\subsection{Dependence on the Feedback Model}\label{sec:model}

Given the sophisticated treatment of the ISM in many current stellar feedback simulations \citep[e.g.][]{Gatto17, Kim18, Colling18, Kannan18}, the simplicity of the feedback model employed in this work may present a source of  uncertainty. As mentioned in Section~\ref{sec:simulations}, we do not attempt to resolve low-temperature ($T < 10^4\,\mathrm{K}$) gas in these simulations due to the additional computational expense, so we cannot model a full three-phase ISM or self-consistent star formation. Nevertheless, we expect our results for gas in the \textit{outflow} to be reasonable, given that higher resolution simulations that do take this lower temperature gas into account find that the majority of outflowing gas is in the cool or hot phase \citep{Kim18}. It then remains to determine whether our simplified model of cluster feedback is sufficient to produce a realistic outflow.

As described in Section~\ref{sec:simulations}, the model employed in this work assumed a small ($< 20$) number of very massive ($M_\mathrm{cl} = 10^7\,\Msun$) clusters were responsible for driving the wind. Our rationale was that in systems like M82, a few of the biggest clusters tend to dominate the energy input in the wind. Additionally, only massive clusters are expected to break out from the disk \citep{MacLow88, Kim17, Fielding18}, so while modeling smaller mass clusters is critical for accurately calculating momentum input in the ISM, we expect them to be less important in determining the physical characteristics of the outflow. Keeping the clusters centralized in the disk also allowed us to analyze these results in a spherical framework, considerably simplifying the analytic description. However, real star-forming galaxies have clusters with a range of masses, and they are often distributed more widely throughout the disk.

To test the extent to which our results depend on the simplifying assumptions made in our feedback model, we have run an additional CGOLS simulation with a more realistic cluster mass function ranging from $M_\mathrm{cl} = 10^4\,\Msun$ to $M_\mathrm{cl} = 5\times10^6\,\Msun$. We additionally allowed these clusters to be distributed throughout the disk, assigning their radial locations such that the integrated star formation rate surface density profile matched the gas density profile. Each cluster was turned ``on" for a more realistic 40 Myr, and the mass and energy injection rates were set according to a Starburst99 single burst population synthesis model. In a sense, this distributed model can be thought of as similar to the clustered model presented here but with a lower net star formation rate, given that a smaller fraction of the ``star formation" is in clusters large enough to break out of the disk and contribute significant energy to driving the outflow.

While a full discussion of the results of that simulation is outside the scope of this paper, we note that the results are qualitatively very similar to those reported here. Both simulations produce multiphase outflows with similar profiles in mass, momentum, and energy outflow rate as a function of phase. The velocities in the hot gas are slightly lower ($v \sim 1000\,\mathrm{km}\,\mathrm{s}^{-1}$) in the distributed cluster simulation, which is expected given the smaller maximum cluster masses, and the net energy outflow rate, while still dominated by the hot gas, is lower by about a factor of 3, implying a lower $\alpha_\mathrm{eff}$. Nevertheless, cool gas is accelerated to similar velocities as those found here, and we observe a similar linear relationship between velocity and scalar mass in the cool gas. Thus, we conclude that the results presented in this work are reasonable, given that quantitative adjustments to our reported quantities would be expected for different star formation rates regardless.

\subsection{Comparison to Previous Work}

Although the CGOLS simulations are unique in their resolution over the volume captured here, several recent numerical studies have addressed similar questions, and a comparison to their conclusions is warranted. These include both global wind simulations \citep{Fielding17b, Vijayan18} as well as higher resolution simulations that focus on patches of the ISM \citep{Li17, Kim17, Kim18, Fielding18, Vasiliev19, Vijayan19}. We are excluding in this discussion zoom-in simulations and cosmological models, as their sub-grid prescriptions for supernova feedback and winds generally do not allow for a hot phase to form, making a comparison less relevant.

Generally speaking, the global disk simulations that have been carried out to-date do not have self-consistent star formation and feedback (nor do the CGOLS simulations). Both \cite{Fielding17b} and \cite{Vijayan18} prescribed a smooth galactic disk, similar to our work here, and then injected discrete supernovae \citep{Fielding17b} or OB associations \citep{Vijayan18} in order to generate winds. Because both methods of feedback lacked the large, powerful clusters employed here, they both found less energetic winds. \cite{Fielding17b} find very low-energy winds, with $\alpha \sim 0.01$, which they attribute to their single supernova injection model. This is consistent with our results, in which clustering reduces the energy losses and increases $\alpha$. Both \cite{Fielding18} and \cite{Vijayan18} find values of the mass loading factor, $\beta$, that are consistent with ours, with $\beta$ ranging from a few percent to approximately 50\% of the star formation rate, and never exceeding it. While \cite{Fielding17b} did not have high enough resolution outside the disk to capture the multiphase nature of the winds, \cite{Vijayan18} showed that their winds had a primarily two-phase structure, similar to those presented in this work. They additionally noted some acceleration of the cool gas in the wind to speeds of $\sim 300\,\mathrm{km}\,\mathrm{s}^{-2}$, but were unable to track the cool gas further due to the constraints of their box size and shorter run time of the simulation. The cool gas acceleration seen is consistent with that presented here. While \cite{Fielding17b} find maximum velocities of a few hundred $\mathrm{km}\,\mathrm{s}^{-1}$, \cite{Vijayan18} find maximum velocities of $800\,\mathrm{km}\,\mathrm{s}^{-1}$ for the hot phase, in contrast to the maximum velocities presented here which exceed $1500\,\mathrm{km}\,\mathrm{s}^{-1}$. This is also consistent with the increased energy of the winds resulting from our maximally clustered configuration, versus the OB-association clustering in \cite{Vijayan18}, and the discrete supernovae in \cite{Fielding17b}.

We now turn our attention to an analysis of work that focused on wind properties using simulations that do not capture a full galactic disk, but rather simulate a smaller region (usually $\sim 1$ kpc) at higher resolution. These studies generally have resolution comparable to the CGOLS simulations, of order a few parsecs, and follow the winds out to distances of several kpc. Although most of the studies focus on Milky Way-like gas surface densities, \cite{Li17} explored higher surface density environments more similar to our disk. In that work, the authors find low mass loading rates ($\beta \sim 0.2$) and moderate energy loading ($\alpha \sim 0.3$), which is consistent with our results. They also find that most of the energy stayed in the hot phase, leading to fast, hot outflows with velocities comparable to those presented here. However, little acceleration of the cooler phases was seen. The authors attribute to this to lower resolution outside the disk (they used AMR to decrease the resolution above and below the midplane). This led to little mixing between the phases, which is consistent with our results. While mostly focused on the physics of superbubble breakout, \cite{Fielding18} also study a higher surface density environment with explicitly-added clusters that reach $10^6\,\Msun$. The results they find for their high surface density, high cluster mass simulations are very consistent with ours, with energy loading factors between 10\% and 50\% that decrease slightly with radius, and mass loading factors of around $0.3$ at their largest radii ($0.5\,\mathrm{kpc}$).

Although they focus on lower surface density environments representative of the solar neighborhood, we also compare our results to those of the TIGRESS model, presented in \cite{Kim18}, and specifically their focus on wind properties in \cite{Vijayan19}, as these simulations represent the state-of-the-art in physical realism, incorporating MHD, self-consistent star formation and stellar feedback, and shearing box periodic boundary conditions. The TIGRESS simulations were carried out in a $1\,\mathrm{kpc}\times1\,\mathrm{kpc}$ box which extended to 4~kpc on either side of the disk. In their analysis, the authors split the gas into the same temperature bins used here, allowing us to compare the results directly (though note that our ``cool" gas is their ``warm"). As in our simulations, the TIGRESS models show that the energy budget of the wind is dominated by the hot phase. At high velocities, the authors find that the acceleration of the cool gas is \textit{not} consistent with a ballistic model, as the velocities are too high, and hence, acceleration due to mixing with the hot phase must be at play. In our CGOLS simulations, the cool gas mass flux appears to increase out to a radius of $\sim4\,\mathrm{kpc}$, after which it decreases. Because this increase is present in all phases, we have attributed the increase primarily to our conical geometry, and as demonstrated in Figure~\ref{fig:outflow_fraction}, we are continuously capturing more of the total outflow out to approximately this radius. In TIGRESS, which features a planar geometry, the authors find a decreasing cool gas mass flux as a function of radius out to 4 kpc, which they attribute to a fountain flow. This is consistent with our findings at larger radii, where the total mass outflow rate decreases slightly. As in our simulations, this indicates that despite the transfer of momentum due to mixing, the \textit{total} cool gas mass flux is not increasing with distance as a result of mixing with the hot phase. 

At small radii, the TIGRESS simulation generally shows a larger loss of energy in the hot phase, with an effective energy loading of $<0.1$ by the time the hot gas reaches the scale height of 400~pc. By contrast, in our model, $\alpha_\mathrm{eff}$ has decreased to 0.5 by 400 pc, and only decreases by a small fraction at larger radii, indicating that our extreme clustering model tends to overestimate the hot gas energy loading. In the case where supernovae are less highly clustered, more remnants are expected to cool without breaking out of the disk, and more interaction between the hot bubbles and disk gas is expected in general, which appears to be the source of most of the energy loss  in the wind. As a result of these increased energy losses, the TIGRESS models show an $\alpha$ of only a few percent at large radii, more than 10 times smaller than that predicted here. Between 1 and 4 kpc, however, the energy flux measured in the hot phase in TIGRESS decreases by less than a factor of 2, which is more consistent with our results at larger radii. Nevertheless (and despite a much lower gas surface density and star formation rate), they find a similar mass loading factor for the hot medium, $\beta_\mathrm{eff} \sim 0.1$.

\subsection{On Cloud Acceleration, Mixing, and Destruction}

In recent years, a number of studies have focused particularly on the survival, acceleration, and mixing of cool gas moving through a hot background (see Section~\ref{sec:introduction} for references). In particular, several studies, \citet{Armillotta16}, \citet{Gritton17}, \citet{Gronke18}, and \citet{Gronke19} predict increasing mass of cool gas clouds as the cool gas mixes with the hot background and the resulting intermediate temperature gas cools and condenses. While we observe evidence of substantial mixing in the CGOLS simulation, and the correspondence between the velocity of the cool gas and its scalar value indicates that the momentum is being transferred via this process, we do not see evidence of the total cool gas mass flux increasing with distance in the wind - rather, we find a decreasing cool gas outflow rate with radius beyond a radius of ~4 kpc. So, what might be the cause of this discrepancy? 

First, we note that even with the unprecendented resolution we have employed for a global model, many of the cool gas clouds in our simulation are still unresolved relative to the criterion for mass growth outlined in \citet{Gronke18}. There, the size required for a cool cloud to grow in mass is estimated as a function of the cloud size, overdensity relative to the background, background pressure, wind mach number, and cloud temperature. Using the numbers relevant for cool clouds in our simulations as measured from the radial profiles in Figures~\ref{fig:profiles_hot} and \ref{fig:profiles_cool}, we find that at small radii ($\sim300\,\mathrm{pc}$), clouds larger than roughly a tenth of a parsec are expected to grow in size and accrete material from the background wind. While we do not capture sizes this small, we do note several large clouds near the base of the outflow in e.g. Figure~\ref{fig:slices_35}, which should be resolved by at least 8 cells per cloud radius, the resolution quoted in \citet{Gronke18} to capture cloud growth. At these radii, the cool gas mass outflow rate is indeed rising (see Figure~\ref{fig:fluxes_35}). Unfortunately, our simulation analysis at these radii is incomplete, since the spatial arrangement of our clusters means gas is still feeding into the cone at these radii - the fact that the total gas mass outflow rate continues to rise until $R\sim3\,\mathrm{kpc}$ means that we cannot distinguish between cool cloud mass growth and increasing total mass outflow rates. However, at larger radii, $R\sim 6\,\mathrm{kpc}$, $\dot{M}$ has plateaued and this should not be a problem. Here, we find that cool clouds with sizes larger than a few parsecs should be growing in mass according to the \citet{Gronke18} criterion. Again, we see clouds with radii larger than this that should be resolved in our simulations, but we find no evidence for increasing cool gas mass outflow rates at these radii, and rather find the exact opposite (mass transfer from the cool phase to hot).

\begin{figure}
    \centering
    \includegraphics[width=0.8\linewidth]{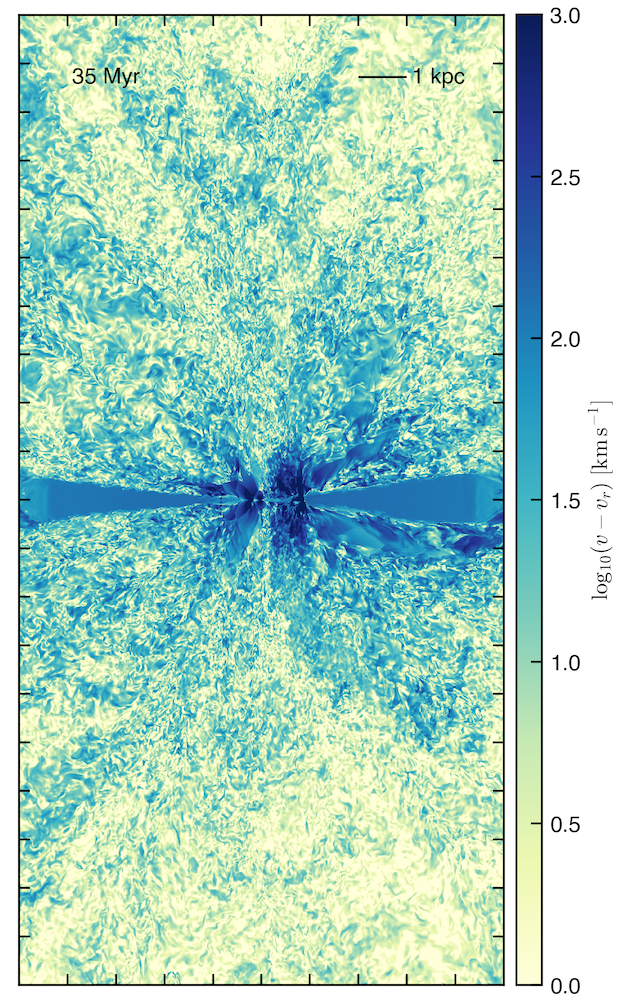}
    \caption{A slice of $\Delta v = v - v_r$ demonstrates the extent to which the wind is turbulent,  rather than a laminar radial flow. Differences between the total velocity and radial velocity are particularly notable at small radii.}
    \label{fig:vvsvr}
\end{figure}

Besides resolution, a primary difference between the clouds in our simulations and those in the simulations discussed above is the nature of the background wind. Whereas the winds are laminar and (in most cases) unchanging in the idealized simulations that see cloud mass growth, our winds are radially expanding and have substantial transverse velocity components that could contribute to disrupting cool clouds and preventing their growth. This is particularly true at small radii, as demonstrated in Figure~\ref{fig:vvsvr}, which shows $\Delta v = v - v_r$ for a slice through the simulation at 35 Myr. As the figure demonstrates, the difference between the total velocity and the radial velocity can be as much as $1000\,\mathrm{km}\,\mathrm{s}^{-1}$, particularly at small radii. As a result, it is not clear whether our simulations actually contradict the results of these more idealized cloud-wind studies, or whether the turbulent background conditions in winds imply that substantially larger cool cloud sizes are needed in order to see cool gas mass growth. In particular, it is entirely plausible that the process outlined in \citep{Gronke18} is happening in our winds, and indeed is the reason for the cool gas acceleration, but the cool gas growth via mixing and cooling is balanced or overpowered by destructive processes owing to the turbulent structure of the background wind.

\section{Summary and Conclusions}\label{sec:conclusions}

In this work, we have presented the fourth simulation in the CGOLS suite of global galactic wind simulations. By employing a unique cluster feedback scheme, we are able to drive energetic hot winds from a high surface density galaxy disk, which give rise to a complex, multiphase outflow. We investigate in detail the properties of this outflow, focusing on separation into two main phases: hot $T > 5\times10^5\,\mathrm{K}$ and cool $T < 2\times 10^4\,\mathrm{K}$ gas. In particular, we find:
\begin{enumerate}
    \item Hot and cool gas co-exist at all radii probed by this simulation, 0-10 kpc. While the cool gas densities as a function of radius are well represented by radial expansion, mass transfer from the cool phase to the hot leads to a shallower density profile than $r^{-2}$ for the hot phase.
    \item Hot and cool gas are not in pressure equilibrium in the wind. The cool gas is under-pressurized by up to a factor of 10 relative to the hot, particularly at small radii where cooling times of clouds are short relative to sound crossing times.
    \item Mixing between hot and cool gas in the wind is an effective way of transferring momentum from one phase to another and occurs at all radii. In cases where the mixed gas has high enough density to be able to cool again, it does so with a higher velocity, leading to a linear relationship between mixed fraction and velocity. This process accelerates cool gas to $> 600\,\mathrm{km}\,\mathrm{s}^{-1}$ by 8 kpc.
    \item The winds produced are highly energetic, with small ($< 60\%$) energy losses relative to the available supernova energy. This likely depends on the degree of clustering in the model employed for feedback. Only a small fraction (~1\%) of the available energy is transferred to kinetic energy of the cool gas. The hot phase dominates both the energy and momentum outflow rates.
    \item All of the above conclusions hold at both star formation rates investigated (5 and 20 $\Msun\,\mathrm{yr}^{-1}$), but the fraction of gas in the cool versus the hot phase is lower for lower star formation rates, as is the radial extent of the cool gas.
\end{enumerate}

Further work remains to be done in comparing the results of these simulations to a framework with a more realistic star formation and supernova feedback model, and ultimately a more realistic three phase ISM. Additionally, comparisons with observables such as absorption lines and covering fractions will further discriminate between these models and others. We will investigate both of these directions in future work.

\acknowledgments
EES thanks Drummond Fielding, Greg Bryan, and other members of the SMAUG CGM collaboration for many helpful discussions that improved the quality of this work, as well as Sean Johnson, Kate Rubin, Gwen Rudie, and Jess Werk, for helping make sense of some the many observations that inspired these simulations. This research used resources of the Oak Ridge Leadership Computing Facility, which is a DOE Office of Science User Facility supported under Contract DE-AC05-00OR22725, via an award of computing time on the Titan system provided by the INCITE program (AST125). EES is supported by NASA through Hubble Fellowship grant \#HF-51397.001-A awarded by the Space Telescope Science Institute, which is operated by the Association of Universities for Research in Astronomy, Inc., for NASA, under contract NAS 5-26555. The work of ECO was partly supported by grant 510940 from the Simons Foundation. 
BER was supported in part by a Maureen and John Hendricks Visiting Professorship at the Institute for Advanced Study, NASA contract NNG16PJ25C and grant 80NSSC18K0563, and NSF award \#1828315.
TAT was supported in part by an IBM Einstein Fellowship from the Institute for Advanced Study, Princeton, and by a Simons Foundation Fellowship during the completion of this work, and acknowledges support from NSF grant \#1516967 and NASA grant 80NSSC18K0526.  

\software{Cholla \citep{Schneider15}; \texttt{matplotlib} \citep{Hunter07}, \texttt{numpy} \citep{VanDerWalt11}, \texttt{hdf5} \citep{hdf5}; Cloudy \citep{Ferland13}, IndeX}

\bibliography{all}

%% Include this line if you are using the \added, \replaced, \deleted
%% commands to see a summary list of all changes at the end of the article.
%\listofchanges

\end{document}